# Reference compositions for bismuth telluride thermoelectric materials for low-temperature power generation

Nirma Kumari,[1] Jaywan Chung,[1] Seunghyun Oh,[1,2] Jeongin Jang,[1] Jongho Park,[1] Ji Hui Son,[1] SuDong Park,[1] and Byungki Ryu,[1,3*]

1) Energy Conversion Research Center, Electrical Materials Division, Korea Electrotechnology Research Institute (KERI), Changwon, 51543, Republic of Korea

2) School of Electronic and Electrical Engineering, Kyungpook National University, Daegu, 41566, Republic of Korea

3) Electric Energy and Material Engineering, School of KERI, University of Science and Technology, Changwon, 51543, Republic of Korea

*Correspondence: byungkiryu@keri.re.kr

# Abstract

Thermoelectric (TE) technology enables direct heat-to-electricity conversion and is gaining attention as a clean, fuel-saving, and carbon-neutral solution for industrial, automotive, and marine applications. Despite nearly a century of research, apart from successes in deep-space power sources and solid-state cooling modules, the industrialization and commercialization of TE power generation remain limited. Since the new millennium, nanostructured bulk materials have accelerated the discovery of new TE systems. However, due to limited access to high-temperature heat sources, energy harvesting still relies almost exclusively on BiTe-based alloys, which are the only system operating stably near room temperature. Although many BiTe-based compositions have been proposed, concerns over reproducibility, reliability, and lifetime continue to hinder industrial adoption. Here, we aim to develop reference BiTe-based thermoelectric materials through data-driven analysis of Starrydata2, the world's largest thermoelectric database. We identify $Bi_{0.46}Sb_{1.54}Te_3$ and $Bi_2Te_{2.7}Se_{0.3}$ as the most frequently studied ternary compositions. These were synthesized using hot pressing and spark-plasma sintering. Thermoelectric properties were evaluated with respect to the processing method and measurement direction. The results align closely with the median of reported data, confirming the representativeness of the selected compositions. We propose these as reference BiTe materials, accompanied by transparent data and validated benchmarks. Their use can support the standardization of TE legs and modules while accelerating performance evaluation and industrial integration. We further estimated the performance of a thermoelectric module made from the reference composition, which gives the power output of over 2.51 W and an efficiency of 3.58% at a temperature difference of 120 K.

## Highlights

- Thermoelectric reference compositions identified for BiTe through data-driven analysis
- $Bi_{0.46}Sb_{1.54}Te_3$ and $Bi_2Te_{2.7}Se_{0.3}$ proposed as p- and n-type reference materials
- The proposed reference compositions synthesized through sintering methods
- Measured results align with the median of literature data, confirming representativeness
- Finite element simulations predict module performance of 2.51 W and 3.58% at $\Delta T$ = 120 K

# 1. Introduction

Thermoelectric (TE) technology enables direct energy conversion between thermal and electrical energy in both directions, offering a promising path toward eco-friendly, fuel-efficient systems and a carbon-neutral society [1,2]. Especially, it is considered suitable for energy harvesting from industrial waste heat, automotive exhaust, and marine engine systems [3,4]. TE modules can also support wireless sensor networks by powering each sensor through small, localized temperature gradients [5]. In marine applications, large diesel engines are significant sources of heat; however, tightening regulations on carbon emissions demand more efficient fuel alternatives [3]. Similarly, in steel manufacturing, extensive thermal waste can be utilized as a valuable energy resource [6]. One of the most successful examples of TE applications is in deep space exploration missions, such as the Voyager I and II probes, where radioisotope thermoelectric generators (RTGs) have provided power for decades [7,8].

A thermoelectric module consists of multiple pairs of p-type and n-type thermoelectric materials, generating electrical power when the module is placed between hot and cold side temperatures ($T_\mathrm{h}$, $T_\mathrm{c}$). Each leg is a thermoelectric element. Under the temperature gradient, an electrical voltage is induced along the leg. When connected electrically in series and thermally in parallel, electrical current flows through the electrical circuit while heat flows across the legs from hot to cold, making the module a solid-state heat engine. Its performance formula [9,10] can be written as

$$\eta \approx \frac{\Delta T}{T_\mathrm{h}} \cdot \frac{\sqrt{1 + zT_\mathrm{mid}} - 1}{\sqrt{1 + zT_\mathrm{mid}} + T_\mathrm{c}/T_\mathrm{h}}. \quad (1)$$

Here, the temperature difference $\Delta T$ is crucial as it bounds the total efficiency, and $zT$ is the dimensionless thermoelectric figure of merit, defined as $zT = \frac{S^2\sigma}{\kappa}T$, which serves as a defining parameter for the material thermoelectric conversion potential. It is determined by a product between the thermoelectric properties: Seebeck coefficient ($S$), electrical conductivity ($\sigma$), thermal conductivity ($\kappa$), and absolute temperature ($T$). With this $zT$ milestone, numerous high-performance TE material systems have been proposed since the early 2000s, expanding beyond BiTe-based materials [11–13].

While solid-state cooling has reached commercial use, TE power generation still faces barriers to industrial adoption due to limited efficiency, reliability concerns, and stability issues [14]. Though many good thermoelectric materials have been suggested but they often depend on high temperature gradients to achieve competitive efficiency [15]. However, in real applications, the usable temperature difference is often reduced due to source dissipation, thermal contacts and interfaces, and system design limitations [15–17]. Even when high temperatures are theoretically available, heat losses during module installation and operation reduce effective temperatures. Even in an ideal system, Peltier heat flow at optimal power may significantly increase the heat current, reducing the effective temperature by 10-30% [18]. With thermal parasitic resistance, efficiency can be halved, while power consumption can be reduced to less than half of its original value. As a result, most practical TE demonstrations, excluding RTG systems, still rely on BiTe-based materials [6,19]. Alternatives, such as Mg-Si and Mg-Sb-based alloys, have been proposed; however, issues of phase stability, defect-induced material degradation, and long-term performance in actual modules remain unresolved, which still makes them far from industrial deployment [20–23].

BiTe-based alloys have been a central issue in thermoelectric research since the 1960s and remain so today. According to Starrydata2, the world's largest thermoelectric database, more than a thousand samples of BiTe-based TE materials have been compiled [24,15,25], as of December, 2021. The composition range is broad, and the reported $zT$ values span 0 to 2.3, consistently showing peak performance around a 300-500 K temperature range. Binary $Bi_2Te_3$ compositions exhibit both p and n-type conduction nature, while the electrical type can be tuned through stoichiometric control and alloying with $Sb_2Te_3$ or $Bi_2Se_3$ [26–28].

Recent studies have extended to quaternary systems, aiming to tune the microstructure, charge,

and phonon transport to achieve high *zT*. However, the n-type BiTe system still suffers from material instability and reliability issues, as is the case with many other TE systems [29]. These concerns limit transparency and comparability in thermoelectric research from an industrial perspective. Furthermore, the chemical space of BiTe alloys is overly broad, and much of the literature focuses on successful high-*zT* cases, which may introduce bias. This lack of clarity presents a serious obstacle to industrial adoption. Hence, a benchmark system for BiTe-based TE power generation is critically needed.

In this study, we aim to define thermoelectric reference compositions within the BiTe alloy family. Just as silicon serves as a foundation in the semiconductor field [30,31], we propose that BiTe should have a clearly defined reference composition for thermoelectric research. A similar study on creating reference material for high-temperature measurement up to 1000 K of the thermoelectric power factor was conducted by Ziolkowski et al. [32]. Such reference material can offer a reproducible baseline for both academic investigations and industrial development. We analyzed composition and TE property distributions of BiTe-based materials across ternary and quaternary compositional spaces using the Starrydata2 database. Among these, $Bi_{0.46}Sb_{1.54}Te_3$ and $Bi_2Te_{2.7}Se_{0.3}$ were identified as the most frequently reported samples for p-type and n-type, respectively. These compositions were synthesized using two widely accepted sintering methods: hot-pressing (HP) and the more non-equilibrium method, spark plasma sintering (SPS). Given the anisotropic nature of $Bi_2Te_3$-based compounds, directional thermoelectric transport was evaluated along both the A-axis and C-axis, where the C-axis is the direction of sample pressing. The measured TE properties align well with a central trend of the literature, supporting the transport property representativeness of the selective compositions. In the spirit of open and transparent science, we have provided full, detailed data on composition, process, sampling direction, and corresponding directional property data. The defined and declared reference compositions for BiTe may serve as a foundation for standardizing thermoelectric legs and modules, ultimately advancing material evaluation methods and accelerating industrial adoption of thermoelectric power generators.

# 2. Results

## 2.1. Data-driven Selection of Reference Compositions for Bismuth Tellurides

### 2.1.1. Database overview

The Starrydata2 dataset includes approximately 43,601 sample property sets from 7,994 publications, as of December, 2021. The raw dataset in this version, sourced from the platform's official GitHub repository, underwent a ***classical filtering*** process to extract a validated subset for TE properties: 13,469 sample datasets from 3,142 unique publications [18,33]. Each sample dataset contains temperature-dependent TE properties measured for $S$, $\sigma$, $\kappa$, and $zT$. It is further divided into two categories based on their Seebeck coefficient: p-type and n-type. Using this divided dataset for p-type and n-type BiTe-based thermoelectric systems, a comprehensive elemental statistical analysis is conducted to map the distribution of co-occurring elements in the ternaries and quaternaries of BiTe. This thorough compositional study provides insights into the compositional landscape of BiTe. The most frequently occurring composition is the mode of the composition data frame, which provides reference representatives for p-type and n-type BiTe systems. For p-type, $Bi_{0.46}Sb_{1.54}Te_3$ serves as the reference composition, and for n-type, $Bi_2Te_{2.7}Se_{0.3}$ is the reference composition. This composition-based statistical analysis for p-type and n-type BiTe systems is detailed in sections 2.1.2 and 2.1.3, respectively.

### 2.1.2. P-type Reference Composition: $Bi_{0.46}Sb_{1.54}Te_3$.

Among 13,469 valid samples, 721 material compositions are p-type BiTe. The occurrence of a third element other than Bi and Te is assessed based on material compositions in which the atomic fraction of the third element in the chemical formula is 10% or greater. Notably, antimony (Sb) is recognized as the most frequently incorporated element, appearing in 511 samples among 721 compositions of the p-type BiTe-based system, as illustrated in **Figure 1(a)**. In addition to Sb, other elements such as Ge, Sn, Se, In, Cu, Tl, Pb, O, Na, Ga, and Ag were also detected; however, their frequency of occurrence is approximately 7 to 20 times lower than that. Based on the high occurrence frequency of the Sb element, BiSbTe is identified as a reference ternary system for the p-type BiTe system.

We used a two-dimensional (2D) Kernel Density Estimation (KDE) method to identify the occurrence frequency densities of third and fourth elements in BiTe materials. KDE is a distribution-free method to estimate the underlying probability density function of a dataset in a continuous 2D space. We employed a Gaussian kernel for its smoothness and differentiability, replacing each data point is replaced with a Gaussian function and summing them to obtain the overall density. The smoothing level was controlled by a bandwidth parameter, which was automatically adjusted based on the size and spread of the dataset to balance resolution and smoothness.

**Figure 1(b)** shows the KDE density map representing the frequency of reported compositions involving Bi and Sb elements in the BiSbTe ternary system, where white indicates low density and blue indicates high density. The highest-density points of the KDE plot, marked by the intersection of dashed blue lines, identifies the most studied composition. We defined it as the reference composition for p-type BiTe. The most frequently observed atomic percentages of Bi and Sb are 9.29 at.% and 30.71 at.%, respectively, corresponding to the composition of $Bi_{0.46}Sb_{1.54}Te_3$.

A general chemical formula for a ternary system based on the p-type ternary composition BiSbTe can be written as $Bi_xSb_yTe_zD_\delta$, where D is a fourth element (or a dopant), and x, y, z, and δ denote the corresponding atomic molar ratios. Using these, the host fraction (***HF***) and dopant fraction (***DF***) are defined as:

$$\boldsymbol{HF} := \frac{(x+y+z)}{(x+y+z+\delta)} \tag{2}$$

$$\boldsymbol{DF} := \frac{\delta}{(x+y+z+\delta)} \tag{3}$$

where Bi, Sb, and Te elements are considered the constituent atoms for the p-type ternary host. The ***DF*** is calculated when ***HF*** $\geq$ 90%

Figure 2(a) shows the ***HF*** for the 721 material compositions of the p-type BiTe system. Their operational hot side temperature ranges from 285 K to 920 K. Along with their hot side temperature of a measured temperature range, the ***HF*** ranges from approximately 20% to 100%. The frequency of data points near the host fraction of 100% and 50% is high. A host fraction close to 100% indicates that the pristine material is based on the $Bi_xSb_yTe_3$ alloys, making it suitable for low-temperature applications (300 K to 650 K). Host fractions closer to 50% indicate that the material base system is different from $Bi_xSb_yTe_3$, possibly PbTe, $AgSbTe_2$, or GeTe, where Bi or Sb is present in a small amount [24,34,35]. These systems are operational in the mid-temperature range (600-900 K), not for the low-temperature range [15]. The frequency of material composition with different atomic ratios of Sb in the ternary system $Bi_xSb_yTe_3$ ($x + y \leq 2$), coming from 692 material compositions, is shown in Figure 2(b), in which y varies between 0 and 2, indicating that the most probable composition based on the highest number of occurrences has y value approximately equal to 1.5.

Compositional statistical analysis of p-type quaternary system ($Bi_xSb_yTe_zD_\delta$) is carried out based on the BiSbTe host compositions, where the ***HF*** is 90% or more. Such a quaternary system is 551 out of 721 compositions of p-type BiTe. The different kinds of observed dopants are Mo, Tb, Pd, Er, Cs, Co, K, Ce, Zn, Ni, Mg, Cr, O, Ba, S, Ti, Fe, Ge, Sn, Ga, Na, Ag, In, Cu, and Se. The number of times a particular dopant is observed in the 551 compositions of the p-type BiTe system is given by its frequency of occurrence or number of occurrences. The number of occurrences of each dopant in the quaternary system is given in Figure 3(a). The element Se has the highest frequency of occurrence among other dopants, with an occurrence frequency of 24. After Se, the frequency of occurrence, more than five, is observed in Cu, In, Ag, Na, Ga, Sn, and Ge, and these elements are written in descending order of their frequency of occurrence, where Cu frequency of occurrence is 20 and Ge has 7. The occurrence frequency of the other dopants, such as Mo, Tb, Pd, Er, Cs, Co, K, Ce, Zn, Ni, Mg, Cr, O, Ba, S, Ti, Fe, is less than 5. Corresponding to each dopant, the molar atomic fraction percentage range is illustrated by the violin plot in **Figure 3(b)**. The width of the violin plot represents the distribution density or frequency of occurrence of each dopant at a given fraction, while the height spans the full range of reported molar atomic percentages for that dopant.

Within each violin plot, the 25th percentile indicates the value below which 25% of the data fall (lower quartile), the 50th percentile marks the median value (the midpoint of the dataset), and the 75th percentile shows the value below which 75% of the data lie (upper quartile). These percentiles provide insight into the central tendency and spread of dopant concentrations. In cases where the number of samples is small, such as for Ti, Mg, Ga, Na, and Cu, multiple percentiles may overlap and appear as a single value.

Among all dopants, Se, Ge, and Sn exhibit a wide distribution, with molar atomic fractions ranging from approximately 0 to 10%. Cs also shows a high dopant fraction (9%), but only a single occurrence is recorded, which can be verified from **Figure 3(a)**. Other frequently occurring elements, such as Cu, In, Ag, Na, and Ga (each with more than 10 reported compositions), generally exhibit a narrower range, with most molar atomic fractions lying between 0% and 2%.

In TE materials, the fourth elements may act as donors or acceptors with ***DF*** less than 1%; fractions significantly above this are considered to contribute to alloying. The high frequency and broad fraction percentile of Se underscore its key role as an alloying element in the quaternary $Bi_xSb_yTe_zD_\delta$ system. At the same time, Cu, Na, Ag, In, and Ga are more likely to function as acceptors due to their lower concentration levels.

### 2.1.3. N-type Reference Composition: $Bi_2Te_{2.7}Se_{0.3}$

Based on the highest number of occurrences of the Se element, the reference ternary system for n-type BiTe is identified as BiTeSe. Among 13,460 valid samples, 510 material compositions are n-type BiTe. In the n-type BiTe system, in addition to Bi and Te, other elements are present in the chemical formula of BiTe. The occurrence of a third element in the n-type BiTe system is estimated based on material composition, in which the atomic fraction percentile of the third element is 10% or greater. All observed elements, along with their respective occurrences in the 510 n-type BiTe material compositions, are shown in **Figure 4(a)**. The element Se exhibits the highest number of occurrences and is present in 248 of 510 material compositions of the n-type BiTe system. In addition to Se, elements such as Cu, Sb, Pb, Mg, In, S, Ge, and Lu are also present in the n-type BiTe system, but their occurrences are less than 50. In the ternary BiTeSe system, a KDE map of the atomic percentages of Te and Se is plotted as depicted in **Figure 4(b)** to obtain an accurate reference composition. The high-density region in the KDE map of Te and Se gives the most frequently utilized compositions. The most frequently observed atomic percentage of the element is 54.09 at.% for Te and 6.06 at.% for Se. This estimation provides a representative reference composition for n-type BiTe as $Bi_2Te_{2.7}Se_{0.3}$ (more precisely, $Bi_2Te_{2.71}Se_{0.29}$).

A general chemical composition of n-type quaternary BiTe systems is given as $Bi_xTe_ySe_zD_\delta$, where the D is the position of the dopant element of the quaternary system, and x, y, z, and δ are the molar atomic ratios of element Bi, Te, Se, and dopant elements, respectively. Based on the maximum number of occurrences of the third element in 510 material compositions of the n-type BiTe system, the host ternary system is considered as BiTeSe. The host and dopant fractions of the $Bi_xTe_ySe_zD_\delta$ system are calculated using **equations 2** and **3**. For the calculation of dopant fraction in the quaternary system $Bi_xTe_ySe_zD_\delta$, material compositions with a host fraction of 0.9 or higher were considered. The host fraction of the n-type BiTe material system, along with its corresponding hot side temperature, is shown in **Figure 5(a)**. The plot indicates that the host fraction varies between 42% and 100%, while the $T_h$ spans from 250 K to 1250 K. A majority of data points fall in a range where host fraction exceeds 90%, and their associated $T_h$ lies between 50 K to 650 K. This observation suggests that most of n-type TE material in this range are primarily based on the ternary system BiTeSe. Out of 510 n-type BiTe material compositions, 507 cases contain Se atom. These are represented as a ternary system with the general formula $Bi_2Te_ySe_z$ (where $y + z \leq 3$). The frequency distribution of Se content z across these compositions is shown in **Figure 5(b)**. The number of occurrences is highest when the Se molar atomic ratio is approximately 0.3. This reinforces the relevance of $Bi_2Te_{2.7}Se_{0.3}$ as a reference composition for n-type BiTe.

Frequency and fraction of dopant, in the quaternary system $Bi_xTe_ySe_zD_\delta$, are shown in **Figures 6(a) and (b)**. In **Figure 6(a)**, the frequency of each dopant is shown. A total of 29 different dopant elements were identified, including Co, K, Al, Zn, Ni, Sn, Mn, cl, Fe, Li, Sm, Na, Ru, Cd, Cs, S, Y, Au, Ce, Ag, Ga, Pb, Ge, Lu, In, Sb and Cu. Among these, Cu appears most frequently with 50 occurrences, followed by Sb with 16 and Indium with 15. The remaining dopants occur in fewer than 10 compositions each, indicating relatively limited exploration in the n-type BiTe system. **Figure 6(b)** displays the corresponding dopant fraction (in atomic molar percent) of each element using a violin plot. This plot gives the distributions of reported doping levels of each element. The 25th, 50th (median), and 75th percentiles are marked within each violin to represent the interquartile range of doping concentrations. Dopants such as Sb, Ge, Pb, S, and Cs exhibit relatively high dopant fractions, reaching approximately 8–10%, suggesting their role as alloying elements in the system.

In contrast, although Cu has the highest occurrence frequency, its dopant fraction remains consistently below 1%, highlighting its use as a low-concentration dopant likely targeting carrier concentration tuning [36]. On the other hand, Sb, despite its lower occurrence compared to Cu, displays the highest dopant fraction, indicating its use as a significant alloying component in the $Bi_xTe_ySe_zD_\delta$ system to form a particles within or outside the matrix [37]. The contrast between Cu's high frequency with low dopant fraction and Sb's moderate frequency with high dopant fraction underscores the distinct functional roles of these elements in thermoelectric performance optimization. Meanwhile, we would like to note that Cu is a very diffusive element in semiconductors and thermoelectric materials [38,39]. Thereby, it should be very careful when using Cu dopants, otherwise composition will change with different temperatures or chemical environmental conditions.

## 2.2. Experimental Methods

### 2.2.1. Material synthesis and sintering techniques

A summary of the reference composition, synthesis techniques, processing, sintering, and sampling techniques employed is provided in **Table 1** and **Figure 7**. After obtaining the reference composition for p-type $Bi_{0.46}Sb_{1.54}Te_3$ and n-type $Bi_2Te_{2.7}Se_{0.3}$ thermoelectric systems through compositional data analysis, we proceeded with material synthesis. The TE properties of the BiTe system are sensitive to the synthesis techniques employed. Therefore, we employed the widely accepted bulk synthesis route, which involves melting the constituent elements. To ensure a homogeneous composition of the precursor material, we performed ball milling. The resulting powders were then consolidated into dense pellets using two well-established techniques in thermoelectrics: HP and SPS. We chose these two different compaction techniques to observe the effect of the processing techniques on thermoelectric properties. The thermoelectric properties of the BiTe-based alloy exhibit significant anisotropy. To quantify the effect of anisotropy, we conducted a directional measurement of TE properties. Studies on the directional TE properties were conducted both parallel and perpendicular to the sample's pressing direction. HP and SPS samples were used for this study. To perform the direction-based evaluation of the TE properties of the reference composition, slicing of the consolidated pellet is required.

Highly pure elements: Bi, Sb, Te, and Se (purity >99.99%) were used in stoichiometric ratios to synthesize the reference compositions for p-type $Bi_{0.46}Sb_{1.54}Te_3$ and n-type $Bi_{0.46}Sb_{1.54}Te_3$. A total of 30 g of each composition was weighed in the stoichiometric ratio and placed in quartz ampules, which were then sealed under an argon atmosphere at a pressure of 0.5 atm. The ampoules were heated in a rocking tube furnace at 800°C for 10 hours to ensure thorough melting. Afterward, the molten material was rapidly quenched in water and subsequently processed for ball milling. Ball milling was conducted in a stainless steel (SS) jar with SS balls of 10 mm diameter, maintaining a material-to-ball weight ratio of 1:15. The milling process was carried out at 300 rpm for one hour. The resulting powder was then sieved through a 325-mesh sieve to ensure that the particle sizes were below 45 µm. The sieved powders were used to fabricate pellets via two distinct techniques: HP and SPS. Both methods were performed at 420°C under a pressure of 50 MPa, with a 5-minute holding time in an argon atmosphere. SPS utilized graphite dies and punches with an inner diameter of 12.7 mm, whereas HP employed tungsten carbide dies and punches. Both tungsten carbide and graphite dies and punches have the same dimensions. The heating profiles for both sintering techniques were carefully controlled to ensure uniform densification and phase stability. The entire synthesis step is depicted in **Figure 7**.

### 2.2.2. Sample Sectioning for directional TE properties measurement

To evaluate the directional TE properties of the HP and SPS samples in both parallel (ǁ) and perpendicular (⊥) directions to the sample pressing direction, it is necessary to cut or slice the consolidated pellet. The samples required for measuring directional TE properties are obtained from a cylindrical pellet measuring 13 mm in height and 12.7 mm in diameter. We sliced it using a diamond wire cutting machine, as shown in **Figure 7(d)**. After slicing, we obtained two rectangular bars (3 mm × 3 mm × 11 mm), a disk (diameter 12.7 mm, thickness 2 mm), and a rectangular slab (2 mm × 8 mm × 8 mm). The brown-colored shape, a rectangular bar, and the disk-shaped pellet (**Figure 7(b)**) are used to measure the TE properties parallel to the pressing direction. The C-axis measurement indicates the measurement parallel to the pressing direction. The turquoise-colored shape, consisting of a rectangular bar and a slab (**Figure 7(d)**), is used for measuring the TE properties perpendicular to the pressing direction. The A-axis measurement indicates the measurement perpendicular to the pressing direction.

### 2.2.3. TE property measurements

Temperature-dependent TE properties, including the Seebeck coefficient, electrical conductivity, and thermal diffusivity, were measured over a temperature range from room temperature (RT) to 300°C in 50°C increments. Thermal diffusivity measurements were performed on a disk and a rectangular slab, and two rectangular slabs were employed to measure the Seebeck coefficient and electrical conductivity simultaneously. The Seebeck coefficient and electrical conductivity were measured using the ULVAC ZEM-3 system in an inert helium atmosphere. Thermal diffusivity was determined using the NETZSCH Laser Flash Analysis LFA 447 system under ambient conditions. Thermal conductivity was calculated using the relation $\kappa = \rho\alpha C_P$, where $\rho$ is mass density determined via Archimedes' principle, $\alpha$ is thermal diffusivity, and $C_P$ is specific heat capacity estimated using the Dulong-Petit limit and assumed to remain constant over the entire temperature range. We used $C_P$ values of 0.19 J/g/K and 0.157 J/g/K for p-type and n-type compositions, respectively. We isolated the phonon or lattice thermal conductivity from the total thermal conductivity ($\kappa_{Ph} = \kappa - \kappa_e$) by computing electronic contribution of thermal conductivity ($\kappa_e = L\sigma T$), where L is the Lorenz number estimated using the single parabolic model under the assumption of acoustic phonon scattering of charge carriers [40]. This approach gives the temperature-dependent $L$ values in the range of $1.55 \times 10^{-8}$ to $1.76 \times 10^{-8}$ $\mathrm{W \cdot \Omega/K^2}$ for p-type and n-type reference compositions. All measurements were repeated to ensure consistency and minimize experimental uncertainties.

## 2.3. TE properties of Reference composition

### 2.3.1. P-Type $Bi_{0.46}Sb_{1.54}Te_3$

The TE properties of the reference p-type composition $Bi_{0.46}Sb_{1.54}Te_3$, including electrical conductivity, Seebeck coefficient, thermal conductivity, phonon or lattice thermal conductivity, power factor ($PF = S^2\sigma$), and the figure of merit, were analyzed as functions of temperature. These properties, as shown in **Figures 8(a-f)**, along with the corresponding data, are presented in **Table 2**. They were measured on sintered samples prepared using two distinct techniques: HP and SPS

Starting with electrical conductivity, its value decreases with increasing temperature, as expected for heavily doped semiconductors. At room temperature, $\sigma$ values differ noticeably between the two sintering techniques. SPS samples along the A-axis and C-axis exhibit $\sigma$ values of 591 S/cm and 535 S/cm, respectively, while HP samples show approximately 10% lower values. This gap in $\sigma$ diminishes as temperature rises, eventually disappearing at around 250°C. Additionally, anisotropy is evident in both HP and SPS samples, with the A-axis samples exhibiting ~10% higher σ values than the C-axis samples. This difference increases slightly with temperature.

The Seebeck coefficient remains positive across all samples, confirming the p-type nature due to the hole-majority charge carriers. In the temperature range near RT to 100°C, $S$ increases with temperature but declines beyond 150 °C due to intrinsic effects. At RT, the SPS A-axis sample exhibits an $S$ value of 245 μV/K, which is 6% lower than that of the HP A-axis sample. Similar to electrical conductivity, the difference in $S$ values between HP and SPS samples decreases as temperature rises. However, a negligible anisotropy is observed in S, indicating consistent carrier density in all directions.

Regarding thermal conductivity, SPS samples exhibit comparable values along the A-axis and C-axis. At RT, the $\kappa$ values in both directions lie in the range of 0.96–0.99 W/mK and show similar values across the entire temperature range. Anisotropy is observed exclusively in HP samples, with the A-axis sample yielding a 6% higher $\kappa$ at RT compared to the C-axis, and this gap increases significantly with increasing temperature, doubling by 300 °C. In terms of phonon thermal conductivity ($\kappa_{Ph}$), the HP A-axis sample displays 6% higher values than the C-axis at RT. This difference increases to 11% at 300 °C. Conversely, in SPS samples, the C-axis exhibits approximately 7% higher $\kappa_{Ph}$ at RT than the A-axis, with the difference diminishing as temperature rises, ultimately equalizing at 250 °C.

The observed anisotropy in $Bi_2Te_3$ thermoelectric materials is attributed to their unique crystal structure, which consists of quintuple layers arranged in a Te(1)-Bi-Te(2)-Bi-Te(1) sequence [41]. These layers are stacked along the Z or C direction and bonded by the weak Van der Waals force, whereas covalent bonds dominate within the quintuple layers. Electron mobility and phonon velocity across the layers are observed to be lower compared to within the layers.

Combining the effects of pressing technique and anisotropy, the *PF* shows that A-axis samples (*PF* is approximately 3.56 $mWm^{-1}K^{-2}$ for both HP and SPS) outperform C-axis samples (*PF* is around 3.16 $mWm^{-1}K^{-2}$) at RT by approximately 9%. This disparity decreases with increasing temperature, and beyond 150 °C, all samples yield similar *PF* values.

Finally, the *zT* demonstrates distinct behaviors for the two sintering techniques. In HP samples, no anisotropy is detected, as both A-axis and C-axis yield identical *zT* values, with a maximum of 1.2 measured at 50 °C. In SPS samples, however, anisotropy is evident; the A-axis *zT* is 13% higher than the C-axis RT, reaching its peak value of 1.2 at 100 °C.

### 2.3.2. N-Type $Bi_2Te_{2.7}Se_{0.3}$

The temperature-dependent TE properties ($\sigma$, $S$, $\kappa$, $PF$, $\kappa_{Ph}$, and $zT$) of the n-type material $Bi_2Te_{2.7}Se_{0.3}$ were systematically studied for A-axis and C-axis samples processed through spark HP and SPS. These properties are presented in **Figure 9 (a-f)**, and the TE properties data are given in **Table 3**. The electrical conductivity decreases with increasing temperature, consistent with the behavior of degenerate semiconductors, as shown in **Figure 9(a)**. At RT, the HP A-Axis sample exhibits a $\sigma$ value of 1089 S/cm, which is approximately 12% higher than the SPS A-axis value of 947 S/cm. Similarly, the HP C-axis sample ($\sigma$ = 959 S/cm) shows 15% higher conductivity than the SPS C-axis ($\sigma$ = 803 S/cm). This trend persists across the measured temperature range, with differences of 8% for the A-axis and 14% for the C-axis at 300 °C. Unlike the p-type $Bi_{0.46}Sb_{1.54}Te_3$ material, where the σ difference vanishes at higher temperatures, the n-type material maintains this disparity.

The negative Seebeck Coefficient values, as illustrated in **Figure 9(b)**, confirm the n-type nature of the material, indicating electrons as the majority carriers. The magnitude of $S$ increases with temperature in the range of RT–150 °C, which is characteristic of degenerate semiconductors. Beyond 150 °C, intrinsic effects dominate, causing a reduction in $S$. At RT, the HP A-axis sample shows an S value of -146 μV/K, 13% lower in magnitude compared to the SPS A-axis sample (-166 μV/K). This difference decreases with rising temperature and becomes negligible at 300 °C. The absence of anisotropy in $S$ across both sintering techniques aligns with observations from p-type $Bi_{0.46}Sb_{1.54}Te_3$.

Thermal Conductivity initially decreases within the RT–150 °C range due to Umklapp phonon scattering and increases beyond 150 °C due to bipolar conduction effects. These effects appear at a slightly higher temperature compared to the p-type counterpart (at 100 °C). At RT, $\kappa$ is 13% higher in HP samples compared to SPS, as shown in **Figure 9(c)**. This $\kappa$ behavior is analogous to trends observed in p-type $Bi_{0.46}Sb_{1.54}Te_3$.

The power factor, illustrated in **Figure 9(d)**, is higher for SPS samples compared to HP samples for both axes. This effect arises from the higher $S$ values in SPS samples, despite their lower $\sigma$ values compared to HP samples. The anisotropy in $PF$ reflects the anisotropy in $\sigma$; however, this difference diminishes as temperature rises. With the convergence of $S$ values at 300 °C, all samples exhibit similar

PF values.

Phonon Thermal Conductivity is shown in **Figure 9(e)**. The $\kappa_{Ph}$ values for the A-axis and C-axis samples are similar between HP and SPS in the RT–150 °C range. However, at temperatures above 150 °C, SPS samples exhibit higher $\kappa_{Ph}$ values compared to HP. Notably, the C-axis samples show unexpectedly higher $\kappa_{Ph}$ values than the A-axis across all temperatures, with approximately 8% difference at RT and a maximum difference of 12% at 150 °C. This observation diverges from the typical layered thermoelectric materials, where a lower $\kappa_{Ph}$ is typically observed along the C-axis.

The figure of merit, presented in **Figure 9(f)**, is influenced by the combined effects of anisotropy and the sintering technique. SPS samples consistently exhibit higher $zT$ values compared to HP samples for both axes, and the A-axis samples yield higher $zT$ values than their C-axis counterparts. The maximum $zT$ value of 0.9 is observed in the SPS A-axis sample at 150 °C, whereas the HP A-axis sample reaches a peak value of 0.8 at the same temperature.

## 2.4. Representativeness and comparison with literature data

### 2.4.1. TE properties distribution

The measured TE properties of the reference compositions, p-type $Bi_{0.46}Sb_{1.54}Te_3$ and n-type $Bi_2Te_{2.7}Se_{0.3}$, are compared with p-type and n-type BiTe systems obtained from Sterrydata2. The comparison utilizes SPS A-axis experimental data for both p- and n-type samples, as they exhibit optimal performance relative to SPS C-axis and hot-pressed samples. Coming to the BiTe literature, TE properties data are used for compositions that define the reference compositions of the p-type and n-type ternary systems.

In **Figures 10 and 11**, the KDE plot of TE properties for BiTe literature data is overlaid with that of the reference compositions for p-type and n-type. These plots collectively highlight the distribution of literature data, enabling a comparative analysis of how the reference compositions perform within the broader thermoelectric landscape of the BiTe system. The thermoelectric performance of p-type literature data lies in the range of 0 to 950 K, while n-type ranges from 0 to 800 K. TE properties of the reference compositions, p-type $Bi_{0.46}Sb_{1.54}Te_3$ and n-type $Bi_2Te_{2.7}Se_{0.3}$, are situated near the high-density region.

## 2.5. Performance prediction

### 2.5.1. Theoretical performance prediction method

The TE energy conversion performance of a conventional TE power generator module (TGM), composed of p-type and n-type reference compositions, was estimated using the finite element method (FEM) simulations based on a simplified 2D uni-couple TGM model. **Figure 12(a)** shows the conventional 200-pair TGM model, in which p-type and n-type TE elements are connected electrically in series and thermally in parallel. Each leg was assumed to have dimensions of 1.4 mm × 1.4 mm × 1.5 mm and includes eight additional layers beyond the thermoelectric material itself. Due to geometrical complexity and memory limitations, the 3D FEM simulations for 200-pair TGM are computationally demanding. Therefore, instead of 3D simulations, a dimensionally reduced 2D simulations were performed for performance prediction.

**Figure 12(b)** shows the 2D uni-couple TGM model. A closed serial electrical circuit can be formed by connecting each leg using Cu electrode with thickness of 0.2 mm. As thermoelectric materials are semiconducting while Cu is metallic, Ni diffusion barrier layers (thickness of 0.008 mm) were inserted at both ends of the legs to prevent chemical interdiffusion induced contamination. The legs were joined with electrodes using 0.09 mm thick solder layers. $Al_2O_3$ substrates (0.85 mm) are used as electrically insulating and thermally spreading layers in direct contact with heat source and sink. Based on this configuration, 2D FEM simulations were conducted. Overall TGM performance of the 200-pair TGM was estimated by multiplying the calculated uni-couple values of voltage ($V$), power ($P$), and heat input current ($Q_h$) by the number of pairs.

For TE properties of the legs, we used experimentally measured thermoelectric properties of p-type and n-type thermoelectric compositions sintered via HP. Note that HP is one of the most used sintering methods for thermoelectric materials. Since BiTe based materials are anisotropic, we chose transport properties along the sample A-Axis direction (see **Table2 and 3**). The temperature-dependent properties were piecewise-linear interpolated within the measured temperature range and extrapolated as constants outside that range.

Here, we consider two models for TGM performance prediction. ***Model A*** represents an idealized case consisting of only the p-type and n-type thermoelectric legs, with thermal and electrical boundary conditions applied directly to the top and bottom surfaces of the legs. ***Model F*** represents a realistic TGM, including all structural components and parasitic electrical and thermal resistances. In ***Model F***, the contact thermal conductivity ($\kappa_c$) between the TGM and heat reservoirs was set to $\kappa_\mathrm{c} = 11{,}274\ \mathrm{W} \cdot \mathrm{m}^{-2} \cdot K^{-1}$ . Interfacial resistances were modelled immediately above and below the thermoelectric materials. The interfacial electrical resistivity and thermal resistivity were set to $\rho_\mathrm{IF} = 10^{-9}\ \Omega \cdot \mathrm{m}^2$ and $\kappa_\mathrm{IF} = 12{,}000\ \ \mathrm{W} \cdot \mathrm{m}^{-2} \cdot \mathrm{K}^{-1}$, respectively. These contact and interfacial resistance were modelled using a finite-thickness bulk model (0.1 mm) [15,18] rather than idealized interfaces.

Multiphysics FEM simulations were performed using COMSOL Multiphysics software [34], solving coupled partial differential equations for thermal conduction, electrical current flux, and thermoelectric effects [42]. Dirichlet thermal boundary conditions (fixed temperature) were applied to the substate surfaces of the 2D uni-couple. For a given electrical current, the voltage difference between two edges of the bottom Cu electrodes wase used to solve the electrical potential distribution. Based on the solved voltage and temperature distribution inside the TGM, the performance of $P$, $Q_h$, and efficiency $\eta = \frac{P}{Q_h}$ was predicted using thermoelectric equations of electrical and thermal current flux. In the FEM simulations, the numbers of mesh elements were 140 for ***Model A*** and 12,299 ***Model F***, while the corresponding degree of freedom were 1,264 and 47,229, respectively.

### 2.5.2. Performance prediction of TGMs

Though high $\Delta T$ is desirable for high power generation in TGMs, achieving a large $\Delta T$ in practice is challenging. Furthermore, in thermoelectric generators (TEGs), the $\Delta T$ applied on the TGM can be smaller than the actual temperature difference between the heat sink and heat source. In addition, the TGM joining processes are technically demanding, as thermoelectric elements can be damaged during module fabrication. Therefore, in this study, we first focused on evaluating the reference performance at a relatively low $\Delta T = 120$ K.

**Figure 13** shows the current (*I*) dependent predicted performance of the 200-pair TGM under $\Delta T = 120$ K assuming hot-side and cold-side temperatures for TGM are 150°C and 30°C. Overall, the TGM performance is lower than that of the ideal material leg-pair performance, highlighting the importance of module geometry optimization as well as high thermoelectric properties. In ***Model F***, the effective temperature differences ($\Delta T_{\text{eff}}$) applied to p-type and n-type thermoelectric elements were reduced due to the TGM components and parasitic thermal resistances. Plane average $\Delta T_{\text{eff}}$ applied thermoelectric materials in ***Model A*** and ***F*** are 120 K and 81.77 K, respectively. The open-circuit voltage ($V_0$) and heat current ($Q_0$) in ***Model F*** were reduced by 18.8% and 19.5%, respectively, when $\Delta T = 120$ K. This reduction can be simply understood from the following integral formula for open circuit conditions. Assuming that average Seebeck coefficient and thermal conductivity are the same, the $V_0$ and $Q_0$ are proportional to the value of $\Delta T_{\text{eff}}$. Thus, the reduction in $Q_0$ shows similar trend to that of $V_0$. When the current increases, the V is linearly reduced owing to internal electrical resistance.

Meanwhile, the P is optimized at a middle of the current range between 0 and the maximum reference current ($I_{\mathrm{ref}}$) where $V(I_{\mathrm{ref}}) = P(I_{\mathrm{ref}}) = 0$. The heat current at hot side ($Q_h$) is monotonically increasing for $0 \le I \le I_{\mathrm{ref}}$. Thus, thermoelectric conversion efficiency is also current-dependent. It was maximized near the current smaller than that of maximum power condition.

**Figure 14** shows the predicted TGM performance characteristics under various $\Delta T = 1 - 300\ K$, assuming a cold-side temperature is 30°C. In case of ***Model A***, the maximum powers are 4.66 W and 14.2 W when $\Delta T = 120$ and 240 K, respectively. As $\Delta T$ increases, the power output also increases scaling approximately $\Delta T^2$ at low temperature differences. However, beyond $\Delta T = 120$ K, power becomes linearly proportional to $\Delta T$. It is speculated that the bipolar effect induced low Seebeck coefficient performance at high temperature is responsible for reduced power at a wide temperature difference. The predicted thermoelectric efficiency of the ideal material pair reaches 4.87% at $\Delta T$ = 120 K, and further to 7.40% at $\Delta T$ = 240 K. When parasitic resistances are included, both voltage and power output are reduced due to thermal and electrical loss at interfaces and contacts. Under this condition, the maximum power ranges from 2.51 to 7.87 W across $\Delta T$ = 120-240 K, which corresponds to maximum power density ($P_{\mathrm{den}}$) from 0.157 W/cm$^2$ to 0.492 W/cm$^2$. Corresponding efficiencies are reduced to 3.58% and 5.58% at $\Delta T$ = 120 and 240 K, respectively. This corresponds to an efficiency loss of approximately 26%. Notably, the power degradation is more severe than the efficiency loss, reaching approximately 46% relative to the ideal material power output.

From the results, we may estimate the low-temperature harvesting performance. For $\Delta T = 1\ K$, the TGM with contact and interfacial resistance is below 1 mW per TGM. However, when $\Delta T = 10\ K$, the power may reach around 20 mW. It even reaches 685 mW per TGM at $\Delta T = 60\ K$. The results indicate the importance of achieving a larger $\Delta T$, otherwise the performance will be negligibly small below mW scale. The currently reported µW-scale [43] might be due to the poor contact resistance between a rigid module surface and a flexible surface.

## 3. Conclusion

By analyzing a large-scale thermoelectric dataset, this work establishes a data-driven method for identifying reference compositions in $Bi_2Te_3$-based systems. The most frequently reported compositions, $Bi_{0.46}Sb_{1.54}Te_3$ for p-type and $Bi_2Te_{2.7}Se_{0.3}$ for n-type, were selected as reference composition candidates. The reference composition was synthesized using two bulk fabrication methods: hot pressing and spark plasma sintering. Directional measurements of thermoelectric properties from samples produced by both synthesis techniques revealed that A-axis SPS samples consistently outperformed other configurations. Crucially, comparisons with literature data using KDE visualization confirmed that these reference compositions reside within the high-density zones of thermoelectric performance metrics. This validates the utility of frequency analysis not just for composition selection but also for setting realistic performance expectations.

To evaluate the practical module-level performance, we estimated the output of the 200-pair TGM using the finite element method, based on measured material properties. The simulation predicted a maximum power output of 2.51 W and a conversion efficiency of 3.58% under a 120 K temperature difference for a realistic module design that includes contact and interfacial resistances. These results highlight the importance of accounting for parasitic effects in module-level performance assessment.

Future work may apply this data-driven framework to other thermoelectric material systems and extend the performance evaluation to various module architectures, supporting standardized benchmarking and accelerating industrial deployment of thermoelectric technologies.

## Acknowledgement

This work was supported by the Korea Institute of Energy Technology Evaluation and Planning (KETEP) grant funded by the Ministry of Trade, Industry, and Energy (MOTIE) (Grant No. 2021202080023D), by the Primary Research Program of KERI through the National Research Council of Science & Technology (NST) funded by the Ministry of Science and ICT (MSIT) (No. 25A01013), and by the National Research Foundation of Korea (NRF) grant funded by the Korea Government (MSIT) (No. 2022M3C1C8093916), the Republic of Korea. ChatGPT 4o was used to improve language and readability with caution.

## Author contribution

**NK:** Investigation (Data), Visualization, Writing, Review.
**JC:** Conceptualization, Investigation (Data), Visualization, Writing, Review, Funding acquisition.
**SO:** Investigation (FEM simulations), Visualization, Writing, Review.
**JJ, JP, JHS:** Investigation (material synthesis, evaluation, validation).
**SDP:** Conceptualization, Supervision, Writing, Review, Funding acquisition.
**BR:** Design, Conceptualization, Supervision, Data preparation, Investigation (Data, FEM, validation), Writing, Review, Funding acquisition.

## Declaration of Interests

The authors declare that they have no conflict of interest.

## Data Availability

The composition data and related codes can be accessed through Figshare.com (link to be updated upon publication; the associated DOI:10.6084/m9.figshare.29506013 will be activated after the publication). The data supporting the findings of this study are available within the article. Further details can be provided by the corresponding author upon reasonable request.

**Table 1:** Summary of reference compositions along with process and sampling methods.

| **Nominal Reference Composition** | **Melting Temp.(°C) and Time (Hours)** | **Pressing Technique** | **SPS & HP Temp. (°C) & Time (Minutes)** | **Sampling for TE Properties measurement** |
|---|---|---|---|---|
| p-type $Bi_{0.46}Sb_{1.54}Te_3$ | 800°C, 8 hrs. | HP and SPS | 420°C, 5 minutes. | ∥ (C-axis) and ⊥ (A-axis) to the pressing direction of HP and SPS |
| n-type $Bi_2Te_{2.7}Se_{0.3}$ | | | | |

**Table 1: Thermoelectric properties data of p-type reference composition $Bi_{0.46}Sb_{1.54}Te_3$**

| p-type, Reference composition $Bi_{0.46}Sb_{1.54}Te_3$ HP A-axis | | | | | | |
|---|---|---|---|---|---|---|
| **Temp. (°C)** | ***σ* (S/cm)** | ***S* (µV/K)** | ***κ* (W/m/K)** | **$\kappa_{Ph}$ (W/m/K)** | **$S^2\sigma$ (mW/m/K$^2$)** | ***zT*** |
| 25 | 541.35 | 257.80 | 0.9458 | 0.6951 | 3.5980 | 1.1342 |
| 50 | 493.07 | 263.43 | 0.9307 | 0.6839 | 3.4217 | 1.1881 |
| 100 | 400.63 | 261.62 | 0.9307 | 0.6989 | 2.7421 | 1.0994 |
| 150 | 346.47 | 242.19 | 0.9623 | 0.7325 | 2.0323 | 0.8937 |
| 200 | 329.87 | 202.76 | 1.0822 | 0.8301 | 1.3562 | 0.5929 |
| 250 | 335.29 | 162.41 | 1.2552 | 0.9583 | 0.8844 | 0.3686 |
| 300 | 351.00 | 126.11 | 1.4446 | 1.0822 | 0.5582 | 0.2215 |
| **p-type, Reference composition $Bi_{0.46}Sb_{1.54}Te_3$ HP C-axis** | | | | | | |
| 25 | 485.85 | 258.90 | 0.8814 | 0.6566 | 3.2567 | 1.1016 |
| 50 | 442.69 | 263.46 | 0.8675 | 0.6459 | 3.0727 | 1.1445 |
| 100 | 359.76 | 263.93 | 0.8574 | 0.6495 | 2.5061 | 1.0906 |
| 150 | 310.73 | 247.25 | 0.8991 | 0.6937 | 1.8995 | 0.8940 |
| 200 | 293.50 | 213.75 | 0.9799 | 0.7578 | 1.3410 | 0.6475 |
| 250 | 296.67 | 175.93 | 1.1113 | 0.8532 | 0.9182 | 0.4323 |
| 300 | 311.55 | 138.68 | 1.2868 | 0.9727 | 0.5992 | 0.2669 |

| p-type, Reference composition $Bi_{0.46}Sb_{1.54}Te_3$ SPS A-axis | | | | | | |
|---|---|---|---|---|---|---|
| **Temp. (°C)** | ***σ* (S/cm)** | ***S* (µV/K)** | ***κ* (W/m/K)** | **$\kappa_{Ph}$ (W/m/K)** | **$S^2\sigma$ (mW/m/K$^2$)** | ***zT*** |
| 25 | 590.86 | 244.78 | 0.9637 | 0.6881 | 3.5402 | 1.0953 |
| 50 | 540.68 | 250.52 | 0.9450 | 0.6725 | 3.3934 | 1.1604 |
| 100 | 439.92 | 255.86 | 0.9238 | 0.6685 | 2.8799 | 1.1633 |
| 150 | 374.62 | 240.09 | 0.9587 | 0.7100 | 2.1595 | 0.9532 |
| 200 | 345.19 | 212.36 | 1.0411 | 0.7795 | 1.5567 | 0.7075 |
| 250 | 340.14 | 175.72 | 1.1921 | 0.8962 | 1.0503 | 0.4609 |
| 300 | 347.67 | 137.19 | 1.3669 | 1.0155 | 0.6544 | 0.2744 |
| **p-type, Reference composition $Bi_{0.46}Sb_{1.54}Te_3$ SPS C-axis** | | | | | | |
| 25 | 534.91 | 243.21 | 0.9949 | 0.7452 | 3.1640 | 0.9482 |
| 50 | 488.25 | 248.38 | 0.9787 | 0.7323 | 3.0120 | 0.9945 |
| 100 | 394.35 | 252.41 | 0.9462 | 0.7170 | 2.5125 | 0.9908 |
| 150 | 335.16 | 243.38 | 0.9575 | 0.7354 | 1.9852 | 0.8774 |
| 200 | 306.15 | 217.40 | 1.0398 | 0.8088 | 1.4469 | 0.6584 |
| 250 | 298.79 | 185.19 | 1.1547 | 0.8976 | 1.0247 | 0.4643 |
| 300 | 297.27 | 148.71 | 1.3107 | 1.0162 | 0.6574 | 0.2875 |

**Table 2:** Thermoelectric properties data of n-type reference composition $Bi_2Te_{2.7}Se_{0.3}$

| n-type, Reference composition $Bi_2Te_{2.7}Se_{0.3}$ HP A-axis | | | | | | |
|---|---|---|---|---|---|---|
| Temp. (°C) | *σ* (S/cm) | *S* (µV/K) | *κ* (W/m/K) | $\kappa_{Ph}$ (W/m/K) | $S^2\sigma$ (mW/m/K$^2$) | *zT* |
| 25 | 1089.46 | -146.34 | 1.2782 | 0.7146 | 2.3330 | 0.5442 |
| 50 | 1040.50 | -150.32 | 1.2429 | 0.6632 | 2.3513 | 0.6113 |
| 100 | 914.60 | -159.34 | 1.2015 | 0.6212 | 2.3222 | 0.7212 |
| 150 | 808.81 | -164.62 | 1.1686 | 0.5911 | 2.1919 | 0.7937 |
| 200 | 728.24 | -164.15 | 1.1723 | 0.5904 | 1.9622 | 0.7920 |
| 250 | 672.85 | -157.80 | 1.2112 | 0.6113 | 1.6755 | 0.7237 |
| 300 | 636.23 | -146.57 | 1.2965 | 0.6638 | 1.3668 | 0.6042 |
| n-type, Reference composition $Bi_2Te_{2.7}Se_{0.3}$ HP C-axis | | | | | | |
| 25 | 958.85 | -147.82 | 1.2563 | 0.7614 | 2.0951 | 0.4972 |
| 50 | 903.77 | -153.11 | 1.2259 | 0.7246 | 2.1187 | 0.5585 |
| 100 | 793.77 | -161.37 | 1.1881 | 0.6860 | 2.0671 | 0.6492 |
| 150 | 701.73 | -166.61 | 1.1553 | 0.6555 | 1.9480 | 0.7135 |
| 200 | 633.88 | -164.64 | 1.1516 | 0.6454 | 1.7183 | 0.7060 |
| 250 | 589.46 | -158.94 | 1.1979 | 0.6731 | 1.4890 | 0.6503 |
| 300 | 568.91 | -145.26 | 1.2612 | 0.6942 | 1.2004 | 0.5455 |

| n-type, Reference composition $Bi_2Te_{2.7}Se_{0.3}$ SPS A-axis | | | | | | |
|---|---|---|---|---|---|---|
| Temp. (°C) | *σ* (S/cm) | *S* (µV/K) | *κ* (W/m/K) | $\kappa_{Ph}$ (W/m/K) | $S^2\sigma$ (mW/m/K$^2$) | *zT* |
| 25 | 947.42 | -165.03 | 1.1715 | 0.6952 | 2.5803 | 0.6567 |
| 50 | 903.67 | -169.81 | 1.1388 | 0.6495 | 2.6057 | 0.7394 |
| 100 | 799.21 | -177.17 | 1.1037 | 0.6086 | 2.5086 | 0.8481 |
| 150 | 711.26 | -180.40 | 1.0819 | 0.5842 | 2.3147 | 0.9053 |
| 200 | 647.43 | -176.55 | 1.1073 | 0.5984 | 2.0180 | 0.8623 |
| 250 | 606.72 | -164.98 | 1.1727 | 0.6373 | 1.6513 | 0.7366 |
| 300 | 585.61 | -146.83 | 1.2818 | 0.6997 | 1.2625 | 0.5645 |
| n-type, Reference composition $Bi_2Te_{2.7}Se_{0.3}$ SPS C-axis | | | | | | |
| 25 | 802.99 | -165.93 | 1.1631 | 0.7598 | 2.2109 | 0.5668 |
| 50 | 766.84 | -170.52 | 1.1473 | 0.7325 | 2.2298 | 0.6281 |
| 100 | 679.13 | -176.57 | 1.1073 | 0.6863 | 2.1174 | 0.7135 |
| 150 | 604.48 | -178.96 | 1.0855 | 0.6618 | 1.9359 | 0.7546 |
| 200 | 552.47 | -174.95 | 1.1049 | 0.6697 | 1.6910 | 0.7241 |
| 250 | 519.58 | -162.35 | 1.1667 | 0.7065 | 1.3695 | 0.6141 |
| 300 | 504.83 | -144.86 | 1.2842 | 0.7808 | 1.0593 | 0.4728 |

**Table 4:** Predicted thermoelectric performance of 200-pair TGM (40 mm × 40 mm) made of reference bismuth telluride compositions when $\Delta T$ = 120 K and 140 K.

| $\Delta T$ | Model | $V_0$ (V) | $Q_0$ (W) | $P_{max}$ (W) | $\eta_{opt}$ |
|---|---|---|---|---|---|
| 120 K | ***Model A*** | 9.95 | 67.45 | 4.66 | 4.87% |
| | ***Model F*** | 8.15 | 54.58 | 2.51 | 3.58% |
| 240 K | ***Model A*** | 18.50 | 140.12 | 14.20 | 7.40% |
| | ***Model F*** | 15.20 | 111.70 | 7.87 | 5.58% |

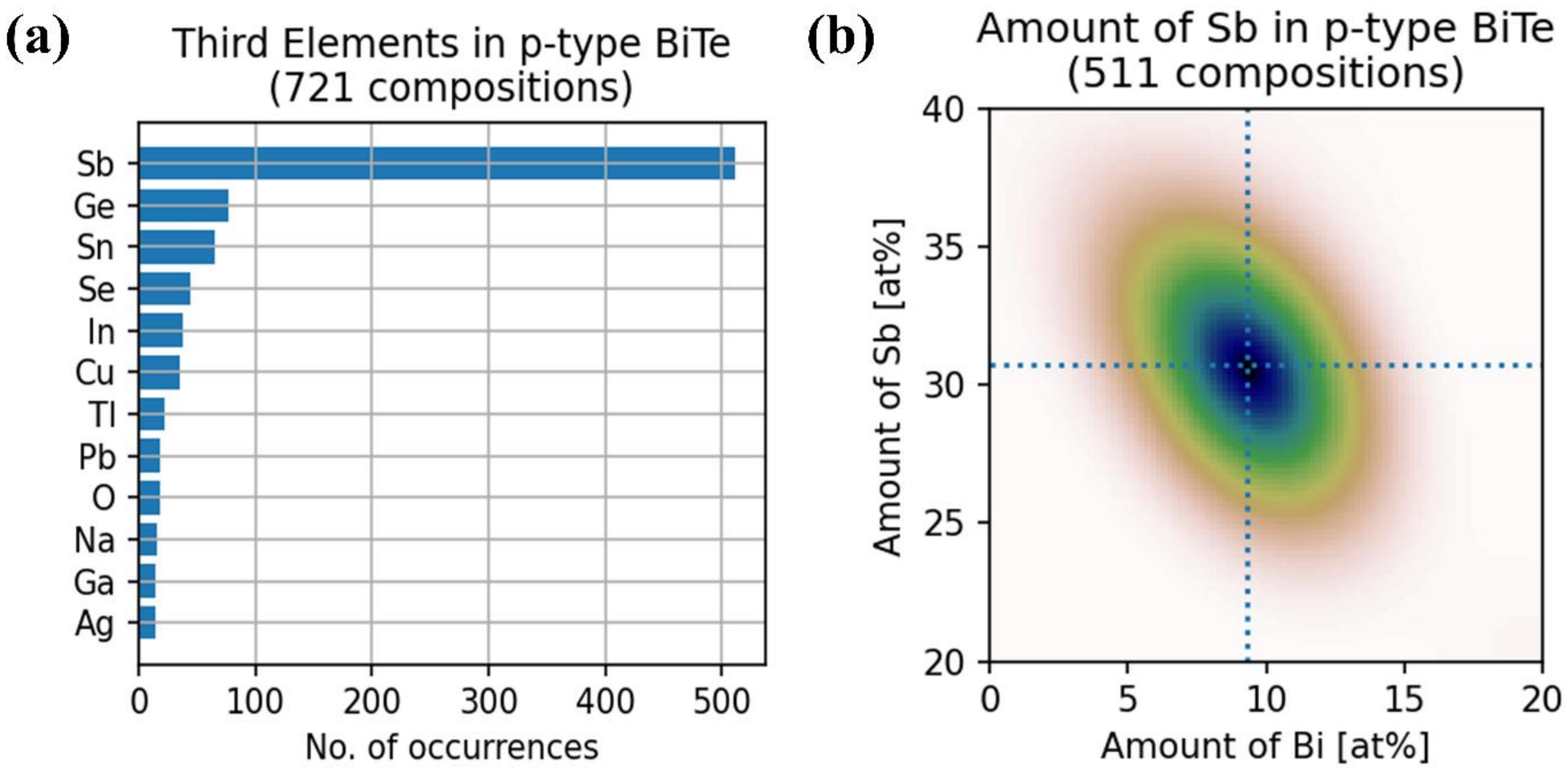

**Figure 1:** (a) Occurrence frequency of the third element in the p-type BiTe system, (b) KDE elemental density map of Bi and Sb of the p-type BiSbTe system.

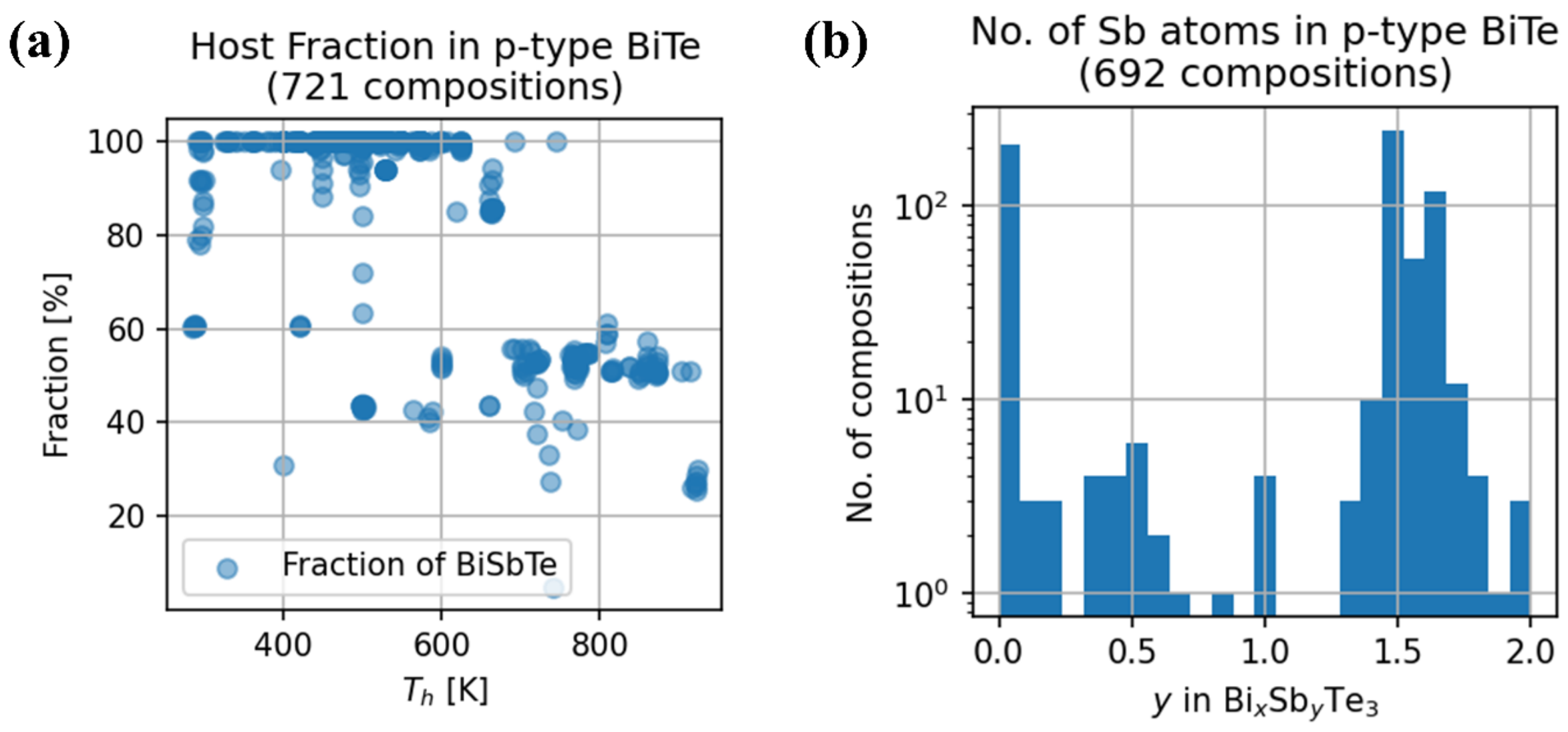

**Figure 2:** (a) Host fraction of p-type BiSbTe system with the respective maximum hot side temperature ($T_h$), (b) Number of compositions with varying molar ratio of Sb elements(y) in $Bi_xSb_yTe_3$.

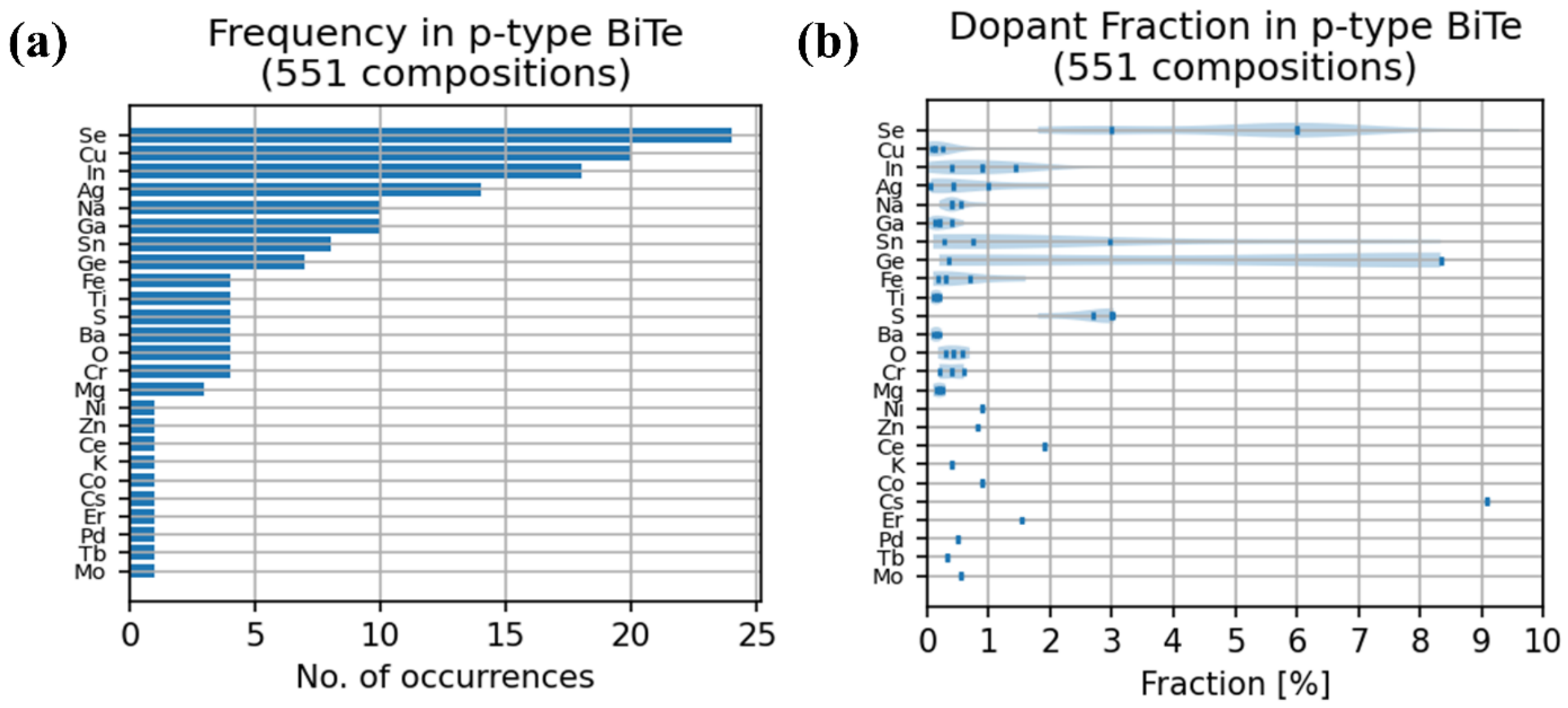

**Figure 3:** (a) Occurrence of frequency of dopant in p-type BiSbTe (b) Dopant fraction in p-type BiSbTe

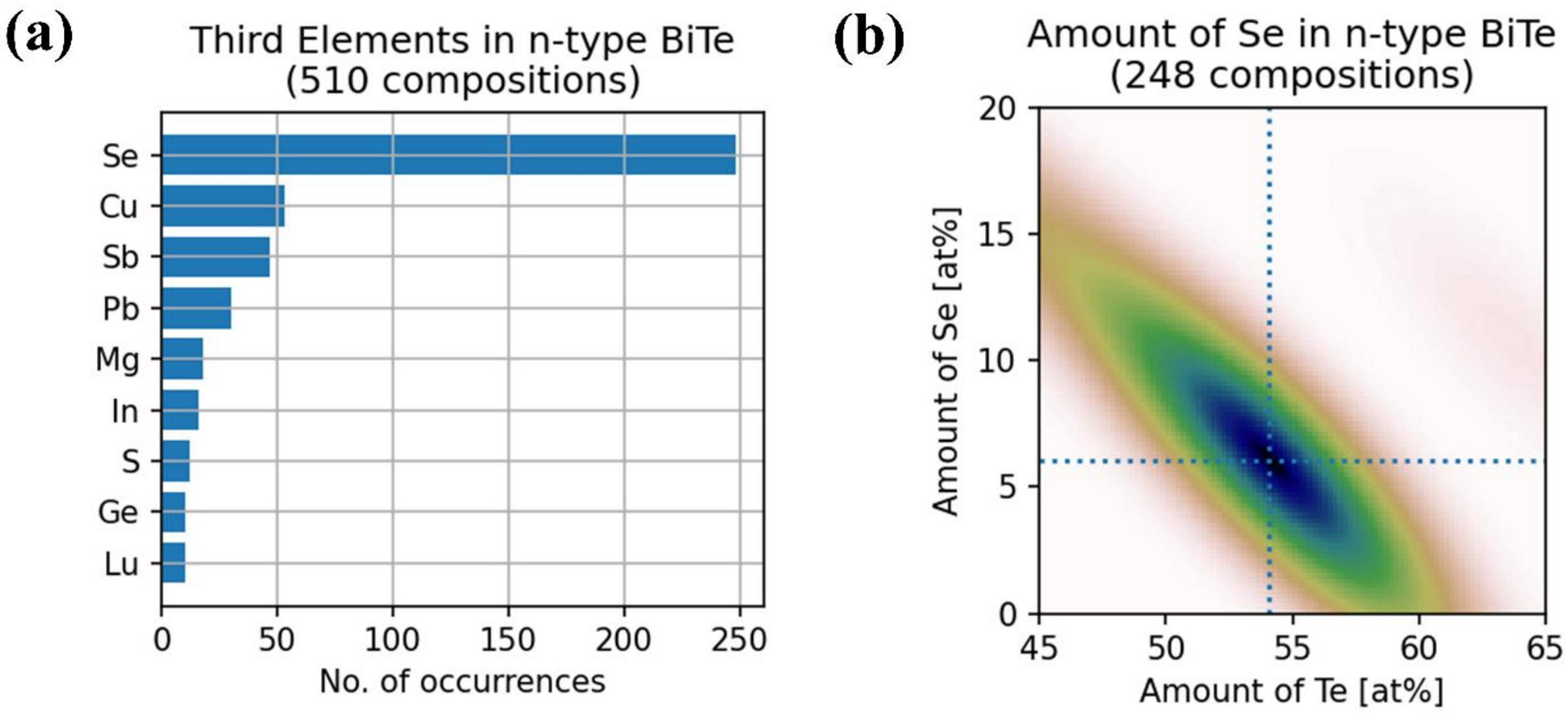

**Figure 4:** (a) Occurrence frequency of the third elements in n-type BiTe system, (b) KDE elemental map of Te and Se of n-type BiTeSe system.

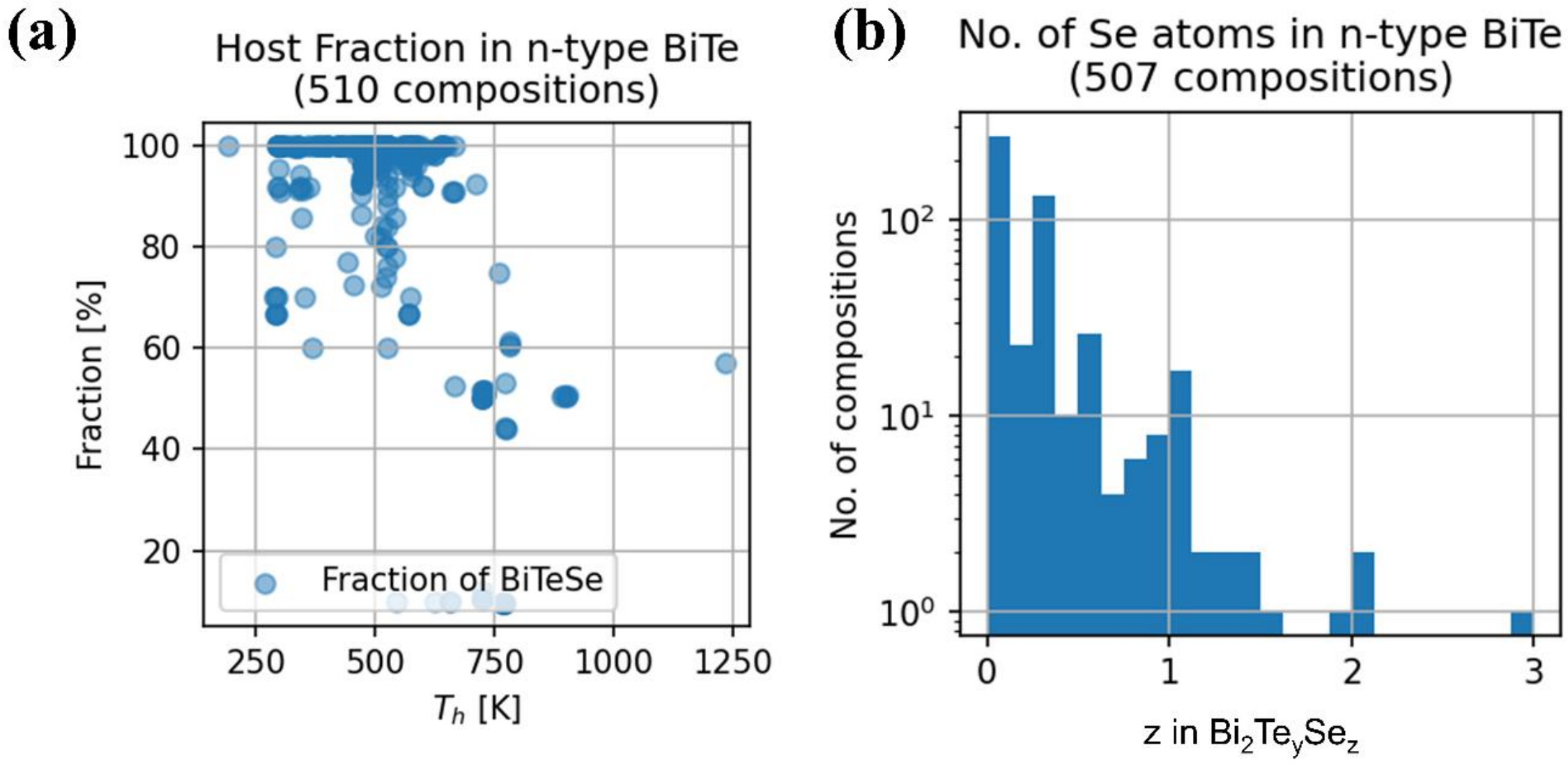

**Figure 5:** (a) Host fraction of BiTeSe system with the respective maximum hot side temperature ($T_h$), (b) Number of compositions with varying molar ratio of Se elements(y) in $Bi_2Te_ySe_z$

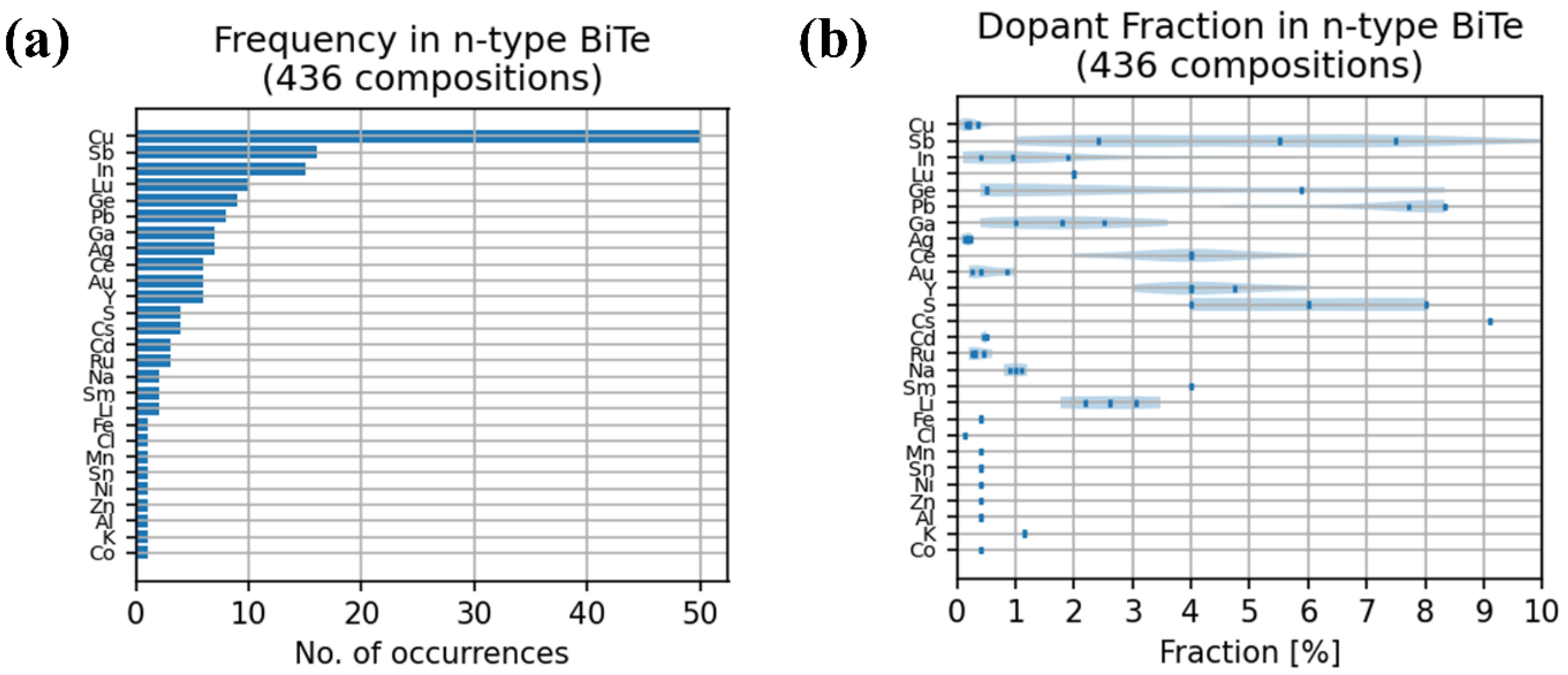

**Figure 6:** (a) Occurrence frequency of dopants in n-type BiTeSe, (b) Dopant fraction in n-type BiTeSe

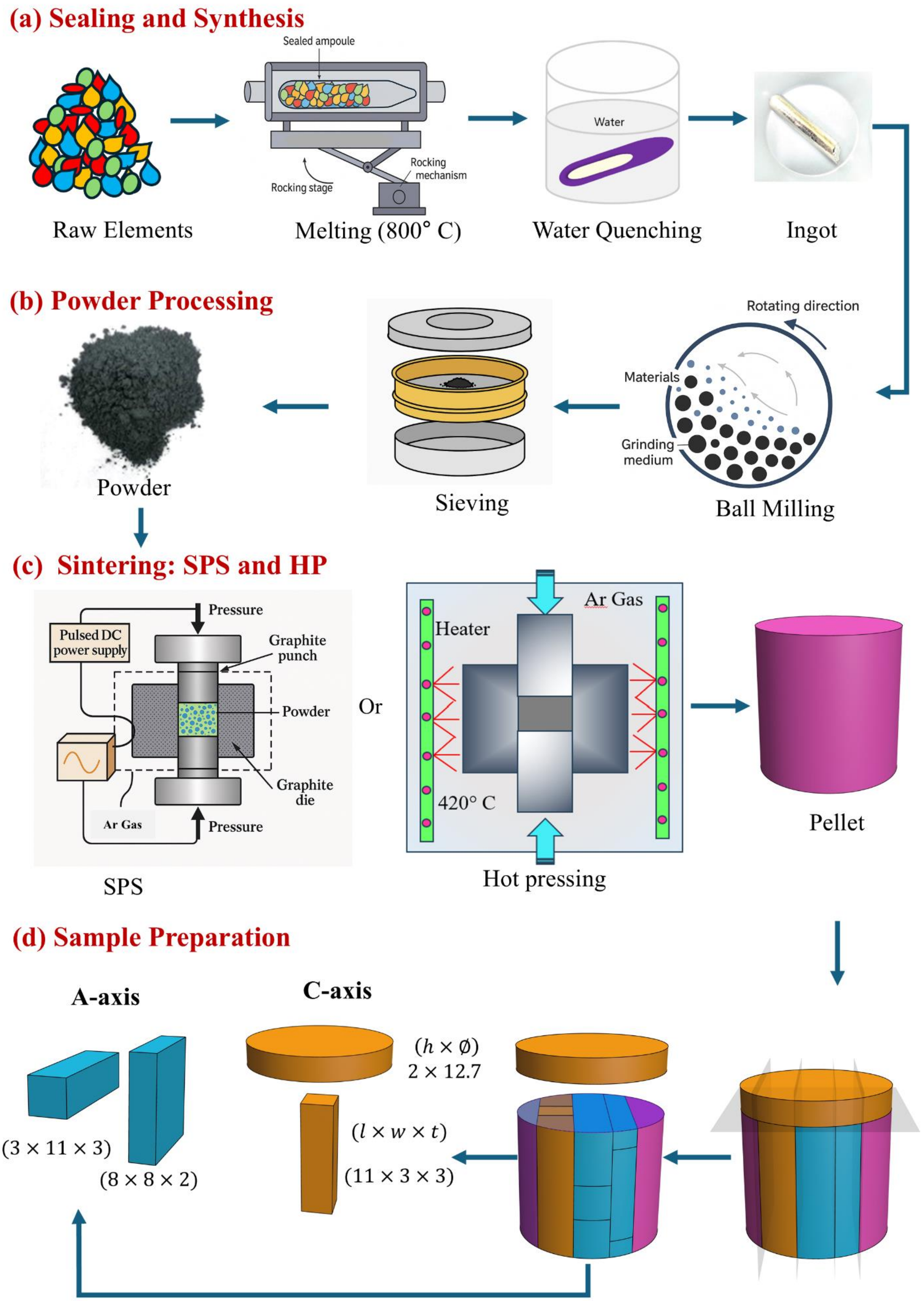

**Figure 7:** Material synthesis, powder processing, sintering, and sample preparation steps.

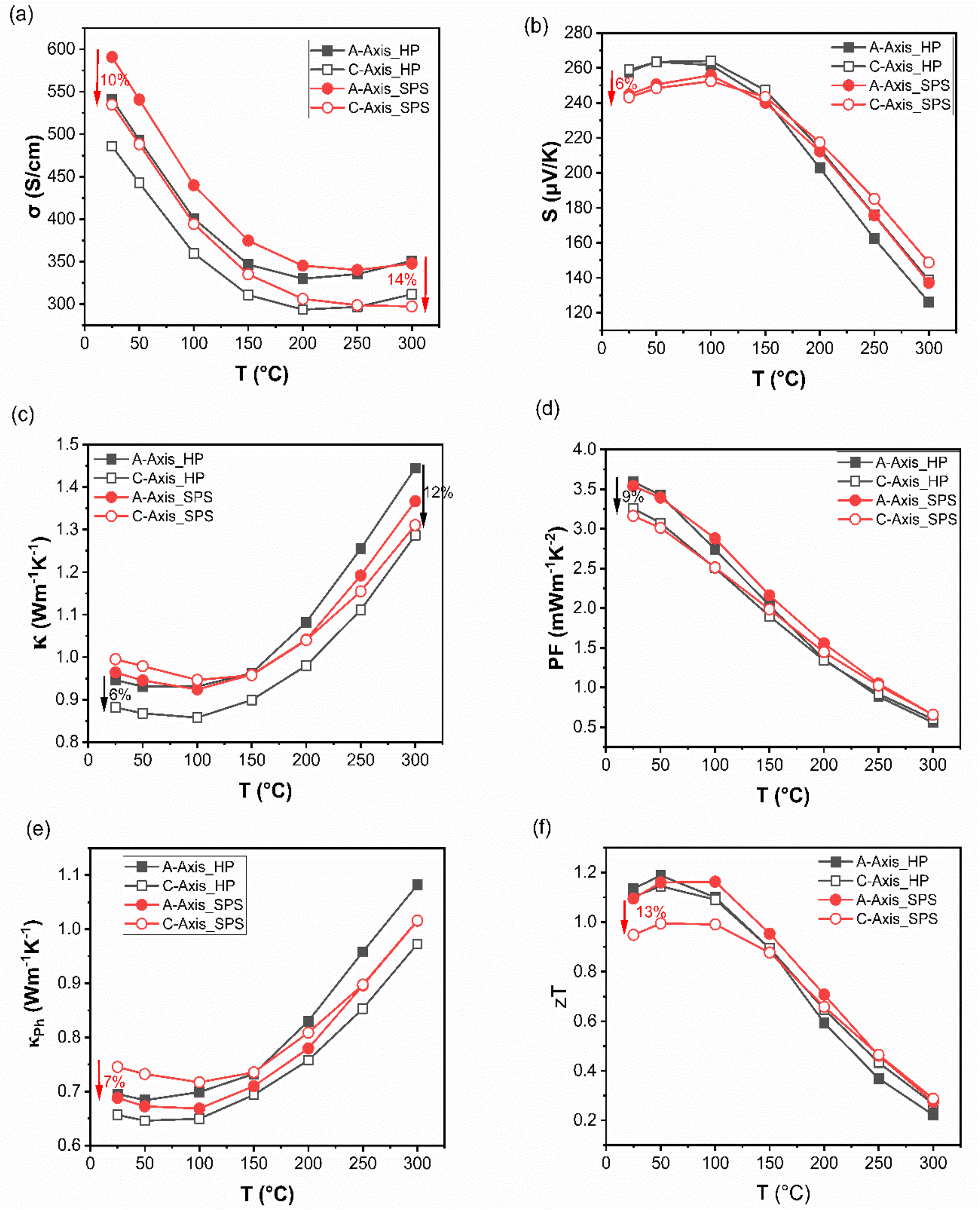

**Figure 8:** Temperature-dependent directional TE properties of p-type TE material $Bi_{0.46}Sb_{1.54}Te_3$, synthesized using HP and SPS. Filled symbol is for the A-axis and the unfilled symbol is C-axis. (a) Electrical conductivity, (b) Seebeck coefficient, (c) total thermal conductivity, (d) power factor, (e) phonon thermal conductivity, and (f) figure of merit.

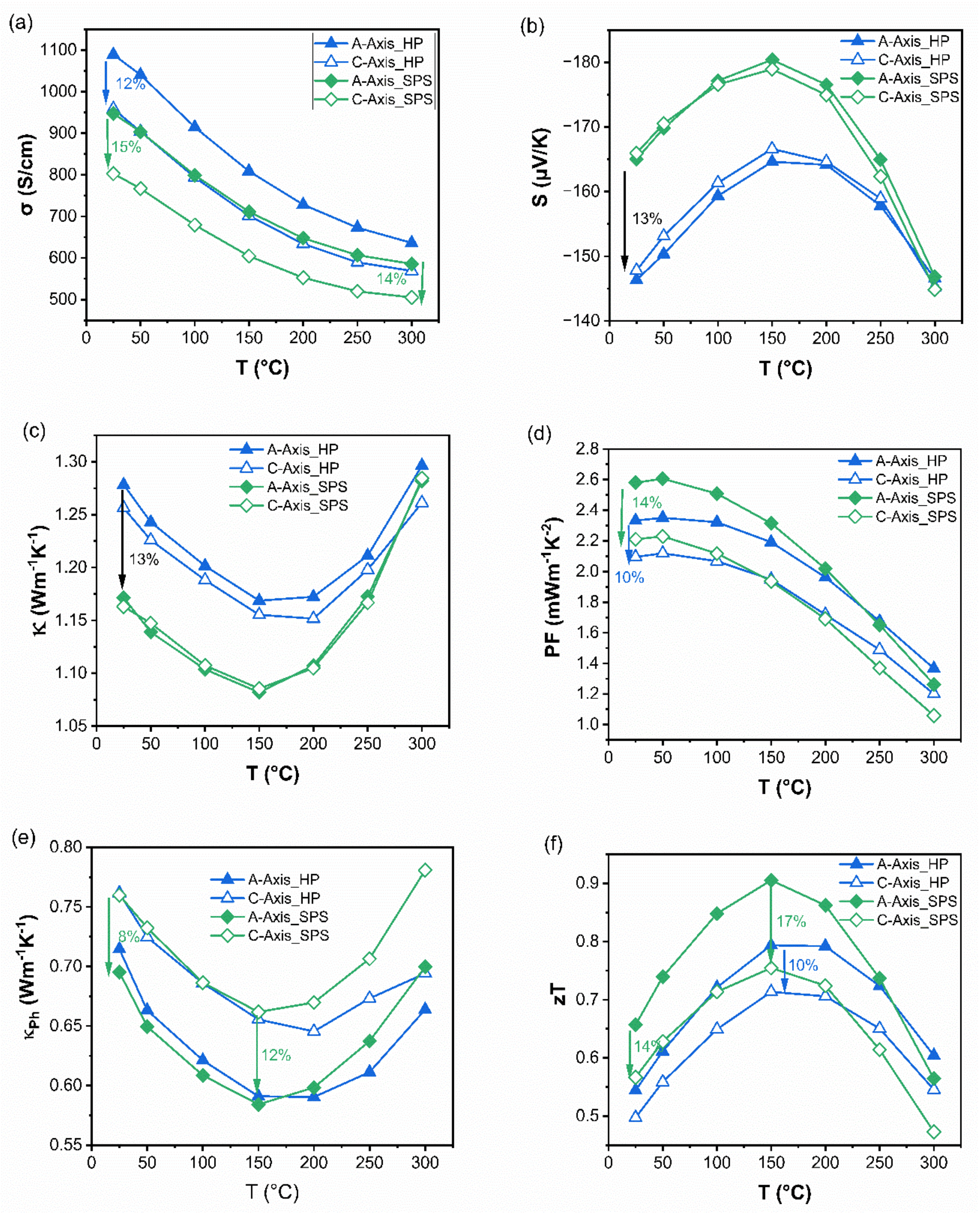

**Figure 9:** Temperature-dependent directional TE properties of n-type TE material $Bi_2Te_{2.7}Se_{0.3}$, synthesized using HP and SPS. Filled symbol is for the A-axis and the unfilled symbol is C-axis. (a) Electrical conductivity, (b) Seebeck coefficient, (c) total thermal conductivity, (d) power factor, (e) phonon thermal conductivity, and (f) figure of merit.

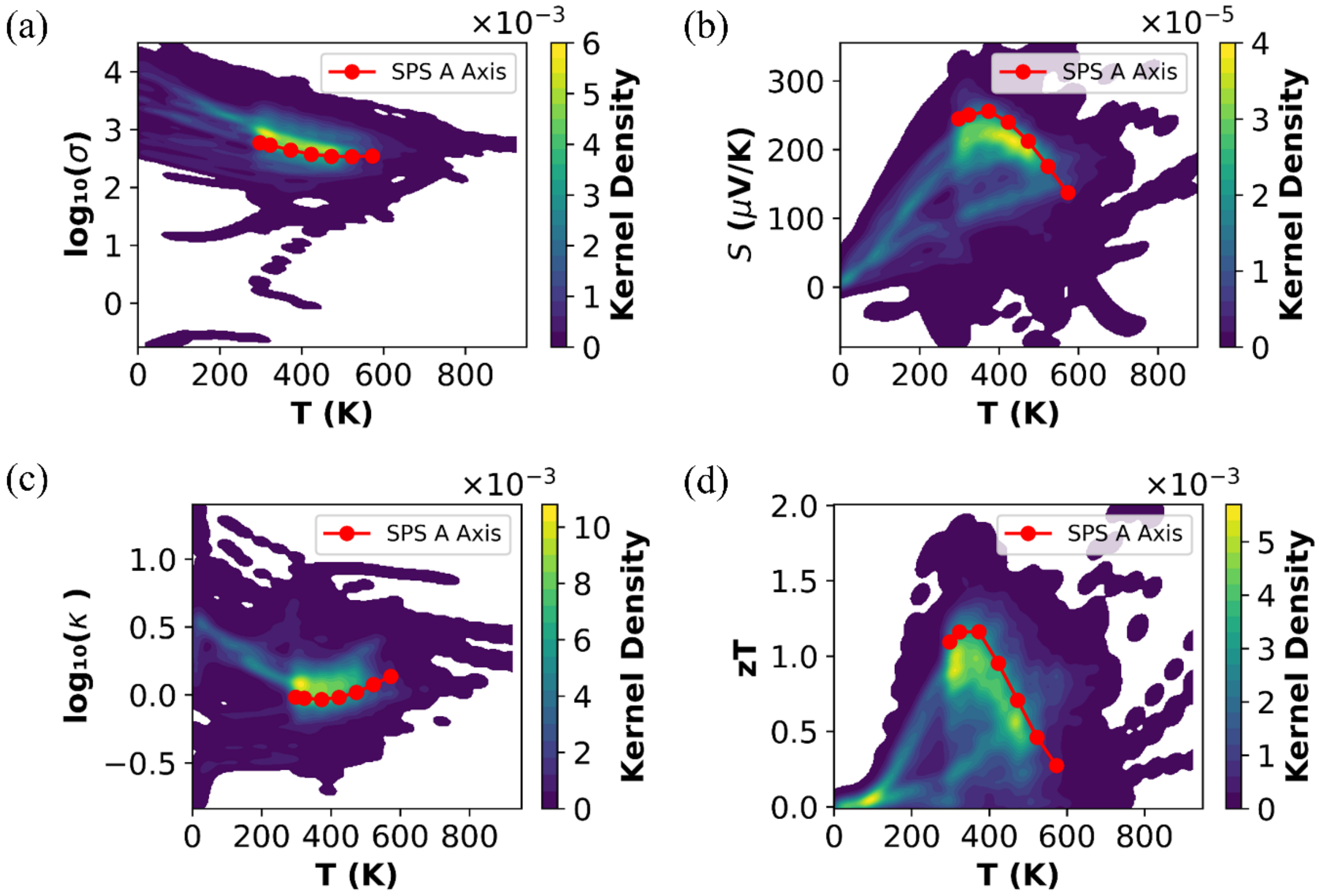

**Figure 10:** (a–d) The KDE plots of thermoelectric properties, electrical conductivity (in the log form σ is normalized by S/cm unit), Seebeck coefficient, thermal conductivity (in the log form *κ* is normalized by W/m/K unit), and figure of merit of p-type BiTe literature data, overlaid with experimental data of p-type reference composition $Bi_{0.46}Sb_{1.54}Te_3$ (red).

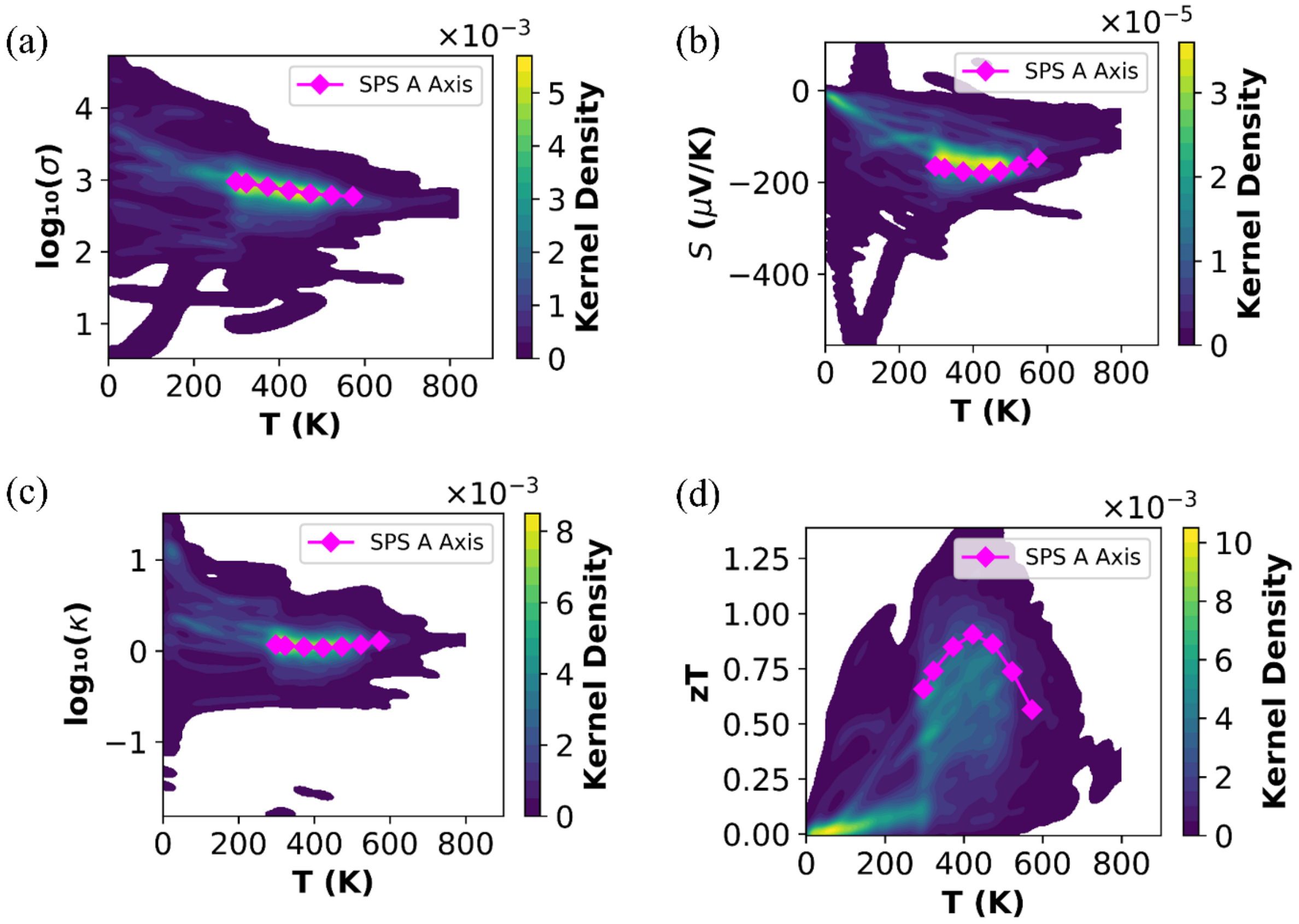

**Figure 11** (a–d) The KDE plots of thermoelectric properties, electrical conductivity(in the log form $\sigma$ is normalized by S/cm unit), Seebeck coefficient, thermal conductivity (in the log form $\kappa$ is normalized by W/m/K unit), and figure of merit of n-type BiTe literature data, overlaid with experimental data of n-type reference composition $Bi_2Te_{2.7}Se_{0.3}$ (magenta).

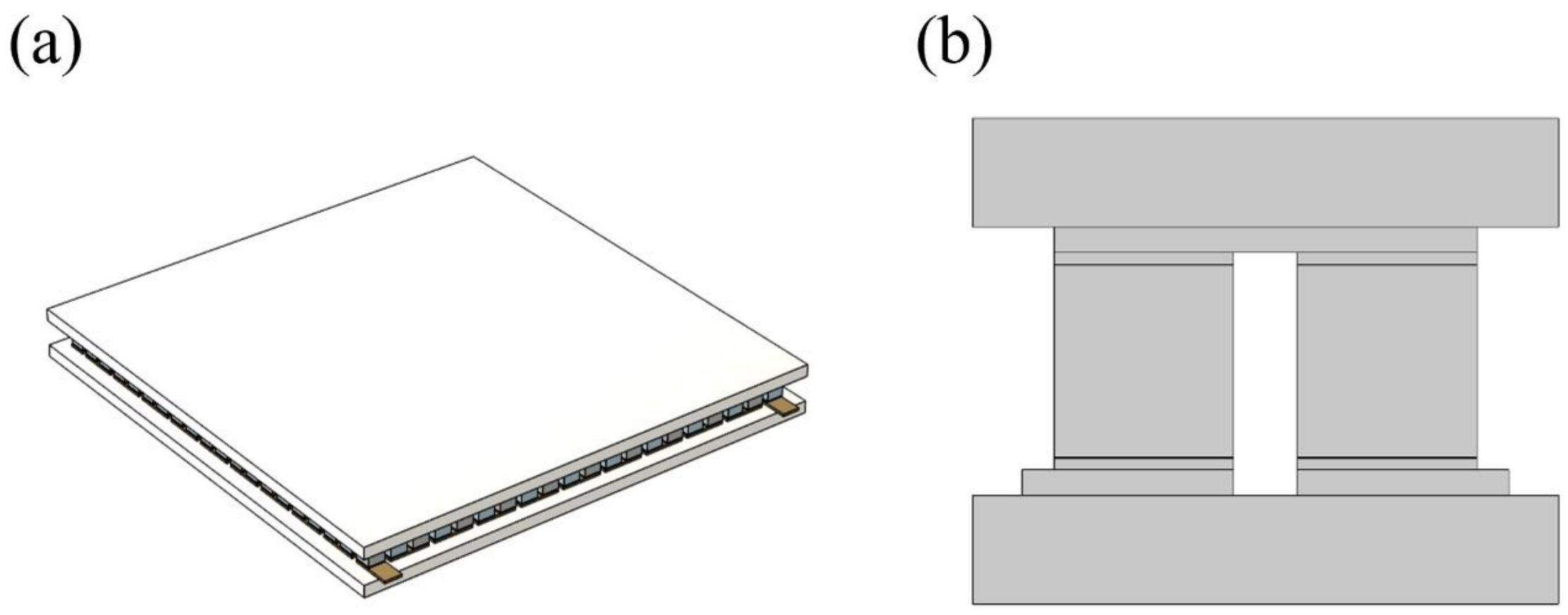

**Figure 12:** (a) Conventional 200-pair TGM model (40 mm × 40 mm × 4.096 mm). (b) 2D uni-couple model for FEM simulations (4.574 mm × 3.796 mm) with a constant depth of 1.4 mm. The performance of the 200-pair TGM can be estimated from the 2D uni-couple models using a simple thermoelectric algebra.

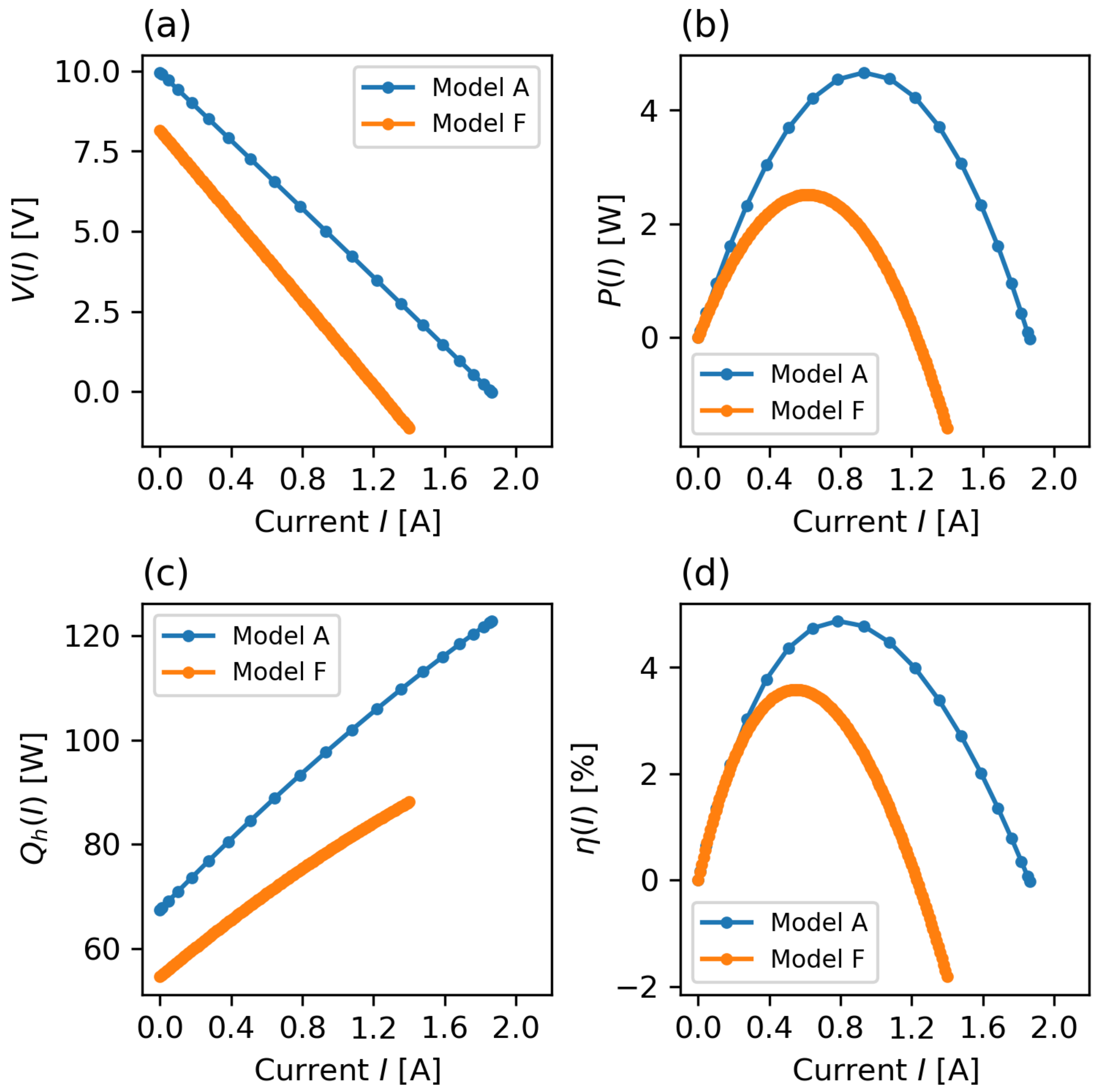

**Figure 13:** Predicted current-dependent module performance characteristics of the 200-pair TGM composed of thermoelectric elements with reference compositions under $\Delta T = 120\ K$. (a) Voltage $V(I)$, (b) power $P(I)$, (c) heat current at the hot side $Q_h(I)$, and (d) thermoelectric conversion efficiency $\eta(I)$. ***Model A*** represents an ideal material pair without any module component and ***Model F*** represents the realistic TGM having full geometries with contact and interfacial resistance.

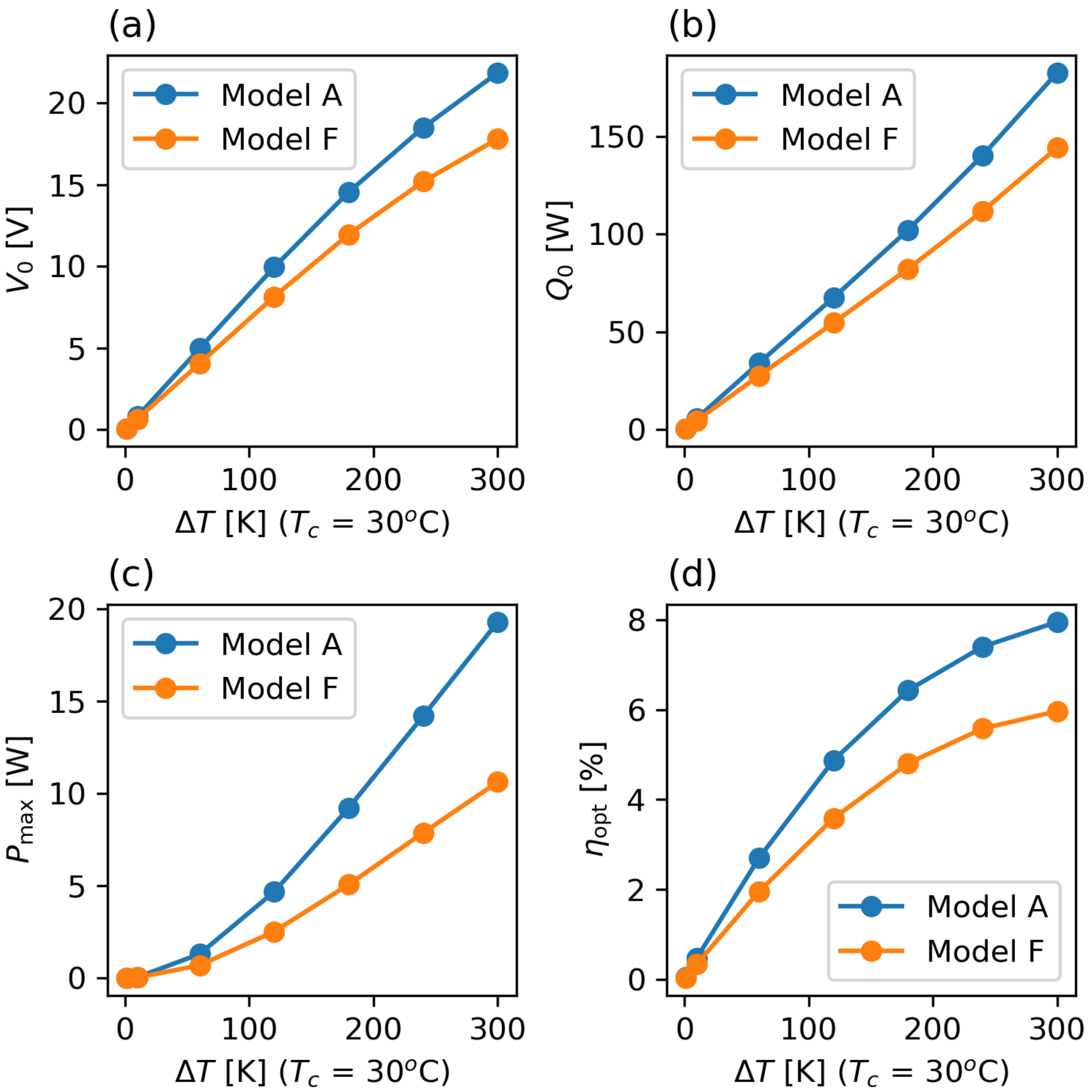

**Figure 14:** Predicted $\Delta T$-dependent module performance characteristics of the 200-pair TGM (40 mm × 40 mm) composed of reference compositions. (a) Open-circuit voltage $V_0$, (b) Open-circuit heat input $Q_0$, (c) maximum power $P_{\max}$, and (d) optimal thermoelectric conversion efficiency $\eta_{\text{opt}}$ over $\Delta T$ range from 1 K to 300 K. The $P_{\max}$ and $\eta_{\text{opt}}$ are determined for each $\Delta T$ by performance optimization.